\documentclass[12pt,showpacs,preprintnumbers]{revtex4}

\usepackage{graphicx}
\usepackage{dcolumn}
\usepackage{bm}
\usepackage{amssymb}
\usepackage{epsfig}    
\usepackage{here}
\def\beq{\begin{equation}}
\def\eeq{\end{equation}}
\def\ba{\begin{array}}
\def\ea{\end{array}}
\def\bea{\begin{eqnarray}}
\def\eea{\end{eqnarray}}

\def\sq2{\sqrt{2}}
\def\End{\end{document}}

\def\Journal#1#2#3#4{{#1} {\bf #2} (#4) #3}

\def\PLB{{\em Phys. Lett.}  B}

\def\PRD{{\em Phys. Rev.} D}

\def\EPC{{\em Euro. Phys. J.} C}

\begin{document}                                                              

\title{$tbW$ vertex in the  \\
Littlest Higgs Model with T-parity}%
\author{%
{F.~Pe\~nu\~nuri\,$^1$},~~{F.~Larios\,$^2$}  
}
\affiliation{%
\vspace*{2mm} 
$^1$
Facultad de  Ingenieria, Universidad Aut\'onoma de
Yucat\'an, A.P. 150, Cordemex, M\'erida, Yucatan, M\'exico. \\
$^2$Departamento de F\'{\i}sica Aplicada,
CINVESTAV-M\'erida, A.P. 73, 97310 M\'erida, Yucat\'an, M\'exico
}

\begin{abstract}
\hspace*{-0.35cm}
A study of the effective $tbW$ vertex is done in the littlest Higgs
model with T parity that includes the one loop induced weak dipole
coefficient $f_{2R}$.  The top's width, the W-boson helicity in the
$t\to bW^+$ decay as well as the t-channel and the s-channel modes
of single top quark production at the LHC are then obtained for the
$tbW$ coupling.  Our calculation is done in the Feynman-'t Hooft
gauge, and we provide details of the analysis, like exact formulas
(to all orders of the expansion variable $v/f$) of masses and
mixing angles of all the particles involved.  Also, a complete and
exact diagonalization (and normalization) of the scalar sector of
the model is made.  

\end{abstract}

\maketitle

\setcounter{footnote}{0}
\renewcommand{\thefootnote}{\arabic{footnote}}

\section{Introduction}

The Top quark plays a major role in the research program of the LHC.
The top is the only quark that decays Weakly before hadronization,
therefore we have an opportunity to study {\it bare} quark properties
like spin, mass and couplings\cite{topreviews,singletops}.
Recent measurements of the single top quark production as well as
the W-helicity in the $t\to bW^+$ decay have been made by the D0 and
CDF groups at the Tevatron and these have (for the first time) set
direct constraints on the effective $tbW$ vertex\cite{tevsingle}.
On the other hand, the high production of top quarks at the LHC
will make it possible to probe directly this vertex down to a few
percent deviation level for the left handed coefficient $f_{1L}$,
and to set limits
of order $10^{-2}$ for $f_{2R}$, and of order $10^{-1}$ for
the right handed $f_{1R}$ and $f_{1L}$\cite{aguilartbw}.
From the theoretical standpoint, observables that depend directly
on the $tbW$ coupling like single top production, the top's width
and the W-helicity in top's decay have been studied in models
beyond the SM like the minimal supersymmetric standard model
\cite{oakes} and the Topcolor assisted
Technicolor (TC2) model\cite{qiao}.

In the Standard Model (SM) the Higgs boson receives large quadratic
divergent corrections from the heavy gauge bosons and from this
fermion.  Models beyond the SM are studied that alleviate this
problem, two important examples are Supersymmetry and Technicolor 
(and TC2)\cite{chivukulalecture}.
Another possible solution is provided by the recently proposed
Little Higgs Models\cite{arkani01,arkani02}
(for a review see Ref.~\cite{littlereviews}).
In these models the quadratic divergent Higgs mass corrections get
canceled at the one loop level via the contribution from certain
(very heavy) partners of the gauge bosons and the top quark
(i.e. the $W_H^{\pm}$ boson, the $Z_H$ boson and the $T$ quark). 
One explicit model has become well known, and it is 
called the ``Littlest Higgs Model'' (LH)\cite{arkani02}.
The LH model is based on a non-linear sigma model of an
$SU(5)/SO(5)$ global symmetry breaking.  It consists of two
$SU(2)\times U(1)$ gauge symmetries that break down to the SM gauge
symmetry at a certain scale $f$.  The phenomenology of the model
deals with heavy partners of the SM gauge bosons, like $W_H^{\pm}$,
$Z_H$ and a heavy {\it photon} $A_H$, as well as a heavy partner
of the top quark $T$\cite{han-logan}.  These heavy partners mix
with the lighter SM gauge bosons and this gives rise to tree level
contributions to precision electroweak observables.  Therefore,
strong constraints have greatly limited the parameter range
of the model (for instance: $f\ge 4$ TeV)\cite{lhlimits}.
A way out of this obstacle is given by implementing a new symmetry
called T-parity, where T-parity even and T-parity odd particles do
not mix\cite{tparity}.  There is one model that is often studied
in the literature; it is based on the previous LH model and is
known as the Littlest Higgs Model with T-parity (LHT)\cite{low}.
Electroweak precision constraints for the LHT model allow the
scale $f$ to be as low as $\sim 500$ GeV\cite{hubisz-precision}.
This model has therefore received more attention recently, with
many phenomenological studies on production and decays of the
new heavy particles\cite{lhtphenomenology} as well as theoretical
studies such as T-parity violation\cite{hill}, top quark induced
vacuum alignment\cite{grinstein}, and two vacuum expectation value
(VEV) scales $f$ in LH models\cite{barcelo}.

In this paper we study the $tbW$ vertex in the context of the
Littlest Higgs Model with T-parity (LHT).  We will often refer
to and will use the notation of Ref.~\cite{belyaev-chen}.
A detailed explanation of the model can be found in
Refs.~\cite{belyaev-chen,hubisz-meade}.  In this work we focus
on the interactions that are relevant to the study of the
effective $tbW$ vertex.

In the literature an expansion in powers of $\epsilon=v/f$
is usually made for the masses and mixing angles derived from
the Lagrangian of the model.  Here, we have obtained the
exact (all powers in $\epsilon\; \equiv v/f$) formulas for
masses and mixings.  Similar expressions have already appeared
in Ref.~\cite{tobe-matsumoto}  and we have found agreement.
Moreover, we provide in detail the
diagonalization procedure of the scalar sector, including the
Goldstone bosons that are eaten by the gauge bosons and that
participate in the one-loop calculation as is done in the
Feynman-t'Hooft gauge.  We provide Feynman rules that are not
found in previous studies of the model.

The next section has the brief presentation of the LHT Lagrangians
(Kinetic and Yukawa)
and the definition of mass eigenstate fields in terms of the original
interaction eigenstates.  Then, in the following section, we will
discuss the effective $tbW$ vertex obtained from tree and one-loop
level contributions.  From this effective vertex we compute some of
the observables associated to the top quark, like the top's decay
width, the W-boson helicity in the $t\to bW^+$ decay, the single
top production process in the two most important modes: the
t-channel and the s-channel.

\section{The Littlest Higgs Model with T-parity.}

The LHT model is based on a non-linear sigma model for an
$SU(5)/SO(5)$ symmetry breaking.  The non-linear $\Sigma$ field
is given as\cite{hubisz-precision}
\bea
\Sigma = e^{2i\Pi/f} \Sigma_0 \;\;,\;\; {\rm with}\;\;\;\;
\Sigma_0 = \left(\begin{array}{ccc}
{\bf 0}_{2\times 2}& {\bf 0}_{2\times 1}& {\bf 1}_{2\times 2}\\
{\bf 0}_{1\times 2}& 1& {\bf 0}_{1\times 2}\\
{\bf 1}_{2\times 2}& {\bf 0}_{2\times 1}& {\bf 0}_{2\times 2}\\
\end{array}\right).\label{sigmafield}
\eea
where $f\sim {\cal O} (1)$ TeV is the symmetry breaking scale known
as the ``pion decay constant''.  The ``pion matrix'' contains a
total of 14 pion fields\cite{hubisz-precision}:

\bea
\Pi\,= \left(\!\!\begin{array}{ccccc}
-\omega^{0}/2- \eta/\sqrt{20} & -\omega^+/\sqrt{2} & -i \pi^+/\sqrt{2} & -i 
\phi^{++} & -i
{\phi^{+}}/{\sqrt{2}} \\
-\omega^-/\sqrt{2} & \omega^0/2- \eta/\sqrt{20}\; & (v+h+i \pi^0)/2 & -i
{\phi^{+}}/{\sqrt{2}} & (-i \phi^0 +\phi_P^0)/{\sqrt{2}} \\
i \pi^-/\sqrt{2} & (v+h-i \pi^0)/2 & \sqrt{4/5} \eta  & -i
\pi^+/\sqrt{2} & (v+h+i \pi^0)/2 \\
i \phi^{--} & i {\phi^{-}}/{\sqrt{2}} & i \pi^-/\sqrt{2} &
-\omega^0/2 - \eta/\sqrt{20} &
- \omega^-/\sqrt{2} \\
i {\phi^{-}}/{\sqrt{2}} & (i \phi^0 +\phi_P^0)/{\sqrt{2}}\; &
(v+h-i \pi^0)/{2} & - \omega^+/\sqrt{2} &\omega^0/2- \eta/\sqrt{20}
\end{array}\!\! \right), \label{14pions}
\eea
Seven of these fields get eaten by the gauge bosons of the model.
The other seven become physical, in particular the $h$ field becomes
the (little) Higgs field whose mass is protected from quadratic
divergencies by the {\it collective symmetry breaking}
mechanism of the Little Higgs model\cite{arkani02}

 An $[SU(2)\times U(1)]^2$ subgroup of the global 
$SU(5)$ symmetry is gauged. The gauged generators have the form
\bea
\label{gaugegenerators}
Q_1^a &=& \left( \begin{array}{ccc} 
\sigma^a/2\; & 0\;\;\; & \;\;\;0\;\;\;\; \\
0 & 0\;\;\; & \;\;\;0\;\;\;\; \\ 
0 & 0\;\;\; & \;\;\;0\;\;\;\;
\end{array}\right), \;\;\;\;  Y_1=
{\rm diag}(3,3,-2,-2,-2)/10\,,\nonumber \\
Q_2^a &=& \left( \begin{array}{ccc} 
\;\;0\;\;\;\; & 0\; & 0\\
\;\;0\;\;\;\; & 0\; & 0 \\
\;\;0\;\;\;\; & 0\; &-\sigma^{a*}/2\end{array} \right),  
\;\;\;Y_2={\rm diag}(2,2,2,-3,-3)/10~.
\eea
The kinetic term for the $\Sigma$ field can be written as
\bea
{\cal L}_{\rm kin} \,=\, \frac{f^2}{8} \mathrm{Tr} D_\mu \Sigma (D^\mu 
\Sigma)^\dagger, \label{kineticlagrangian}
\eea
where
\begin{equation}
D_\mu \Sigma =
\partial_\mu \Sigma - i \sum_j \left[ g_j W_j^a (Q_j^a \Sigma +
\Sigma Q_j^{aT} )+ g'_j B_j( Y_j \Sigma + \Sigma Y_j)\right]\,,
\label{sigmacovariantderiv}
\end{equation}
with $j=1,2$.  Here, $B_j$ and $W^a_j$ are the $U(1)_j$ and
$SU(2)_j$ gauge fields, respectively, and $g^\prime_j$ and $g_j$ are the 
corresponding coupling constants. The vev $\Sigma_0$ breaks the extended 
gauge group $\left[ SU(2)
\times U(1) \right]^2$ down to the diagonal subgroup, which is identified 
with the standard model electroweak group $SU(2)_L \times U(1)_Y$. 

 The field $H$ has the appropriate quantum numbers to be identified
with the SM Higgs; after electroweak symmetry breaking (EWSB), 
it can be decomposed as $H=(-i\pi^+, \frac{v+h+i\pi^0}{\sqrt{2}})^T$,
where $v=246$ GeV is the EWSB scale.

The Lagrangian in Eq.~(\ref{kineticlagrangian}) is invariant under 
T-parity provided that $g_1=g_2\;(\equiv \sqrt{2}g)$ and 
$g^\prime_1=g^\prime_2 \;(\equiv \sqrt{2}g^\prime)$.
The T-parity gauge boson eigenstates (before EWSB) have the simple form,
$W_\pm = 
(W_1 \pm W_2)/\sqrt{2}$, $B_\pm =  (B_1 \pm B_2)/\sqrt{2}$, where $W_+$ and 
$B_+$ are the standard model gauge bosons and are T-even, whereas $W_-$ and 
$B_-$ are the additional, heavy, T-odd states.  From now on we will denote
$W_-$ and $B_-$ as $W_H$ and $B_H$, whereas $W_+$ and and $B_+$ will
be written simply as $W$ and $B$.  After EWSB, the T-even neutral states
$W^3$ and $B$ mix to produce the SM $Z$ boson and the photon.
Since they do not mix with the heavy T-odd states, the Weinberg
angle is given by the usual SM relation, $\tan\theta_w=g^\prime/g$, and
no corrections to precision electroweak observables occur at tree level.
The mixing for the neutral gauge boson mass eigenstates is written as 
\bea
\label{gaugebosonmixing}
\left( \begin{array}{c} Z \\
A \end{array} \right)
= \left( \begin{array}{cc} c_w & -s_w \\
s_w & c_w \end{array} \right) 
\left( \begin{array}{c} W^3 \\
B \end{array} \right) \;\; ,\;\;\;\;
\left( \begin{array}{c} Z_H \\
A_H \end{array} \right)
= \left( \begin{array}{cc} c_H & s_H \\
-s_H & c_H \end{array} \right) 
\left( \begin{array}{c} W_H^3 \\
B_H \end{array} \right) \; ,
\eea
where $s_w = \sin{\theta_w} \; \simeq \sqrt{0.223}$ refers to the SM
mixing angle, and  $s_H = \sin{\theta_H}$ refers to the heavy boson
mixing angle.  Tan$\theta_H$ ($\equiv t_H$) must satisfy the equation:
\bea
t_H^2 + 2 a_H t_H - 1 \;=\; 0 \; ,\;\;\; {\rm with}\;\;\; 
a_H \;=\; 4 (\frac{1}{t_w}-\frac{t_w}{5})/s_v^2\;-\;
(\frac{1}{t_w}-t_w)/2\; , \label{thequation}
\eea
where $t_w = s_w/c_w \;\simeq 0.536$ and
\bea
s_v  \; \equiv \; \sin{\sqrt{2}\epsilon} \;\;\;\; ,\;\;
c_v  \; \equiv \; \cos{\sqrt{2}\epsilon}
\;\;\;\;\; {\rm with} \;\; \epsilon \equiv \frac{v}{f}\; .
\label{epsilon}
\eea

With this value of $t_w$ we obtain
\bea
\tan{\theta_H} \; =\; \sqrt{a_H^2+1}-a_H \;\; \simeq \;
0.142 \epsilon^2 - 0.068 \epsilon^4
\eea

The gauge boson mass terms obtained from Eq.~(\ref{kineticlagrangian})
are as follows:
\bea
M^2_W &=& (\frac{fg}{2})^2 (1-c_v) \nonumber \\
M^2_Z &=& (\frac{fg}{2c_w})^2 (1-c_v) \nonumber \\
M^2_{W_H} &=& (\frac{fg}{2})^2 (3+c_v) \nonumber \\
M^2_{Z_H} &=& (fg)^2 \;(\; 1-\frac{s^2_v}{8}+\delta_{H}\; ) 
\label{bosonmasses} \\
M^2_{A_H} &=& (fg)^2 \;( \; t^2_w (\frac{1}{5}-\frac{s^2_v}{8}) - 
\delta_H \;) \nonumber \\
\delta_{H} &\equiv& \frac{t_H c^2_H}{8} \left(
2 s^2_v t_w \;+ \; t_H [ \; -8+t^2_w (\frac{8}{5} -
s^2_v) +s^2_v \; ] \right) \nonumber 
\eea
Similar expressions are given in Ref.~\cite{tobe-matsumoto};
our formulas are presented differently but agree with theirs
(our mixing angle $\theta_H$ differs in sign).

\subsection{The Yukawa couplings for quarks}

The Little Higgs model introduces a heavy partner of the
top quark (the heavy Top quark) with the purpose of
cancelling out the one loop quadratic divergent radiative
corrections to the Higgs mass\cite{arkani02}.  When
T-parity is implemented in the fermion sector of the model
we require the existence of mirror partners for each of
the original fermions.  This means that for the third family
we have, in addition to the usual bottom and top quarks,
the mirror bottom, the mirror top; as well as the heavy
Top quark with its own mirror quark.

The T-parity invariant Yukawa Lagrangian of the LHT model
is separated into four parts that generate masses for
mirror quarks, down type quarks, first two generations of
up type quarks and finally the top quark and its heavy partner.
It is the latter that is defined in such a way that the
top quark quadratic radiative corrections to the Higgs mass
are canceled.  We are only interested in the Yukawa Lagrangian
of the third family.  A presentation that includes first two
families and the corresponding Cabibbo-Kobayashi-Maskawa
mixing can be found in
Ref.~\cite{buras1}.  For the purpose of our work we consider only
the mirror and top quark Yukawa Lagrangians\cite{belyaev-chen}:
\bea
{\cal L}_{\kappa} &=& -\kappa f \left( \bar \Psi_2
\xi \Psi_c \; +\; \bar \Psi_1 \Sigma_0 \Omega \xi^\dagger
\Omega \Psi_c   \right) \; +\; {\rm h.c.}
\nonumber \\
{\cal L}_{down} &=& \frac{i\lambda_d}{2\sq2}
f \epsilon_{ij} \epsilon_{xyz} \left[ \bar \Psi^{\prime}_2
\Sigma_{iy} \Sigma_{jz} X - \left( \bar \Psi^{\prime}_1
\Sigma_0 \right)_x \tilde \Sigma_{iy} \tilde \Sigma_{jz}
\tilde X \right] d^+_R
\nonumber \\
{\cal L}_{t} &=& -\frac{\lambda_1 f}{2\sq2} 
\epsilon_{ijk} \epsilon_{xy} \left[ (\bar Q_1)_i \Sigma_{jx}
 \Sigma_{ky} \; -\; (\bar Q_2 \Sigma_0)_i \tilde \Sigma_{jx} 
\tilde \Sigma_{ky}   \right]  u^+_{R}\nonumber \\
&-& \lambda_2 f \left( \bar U_{L_1}  U_{R_1}
+ \bar U_{L_2}  U_{R_2} \right)
\;\; +\; {\rm h.c.} \; ,
\label{yukawas}
\eea
where $\xi \equiv e^{i\Pi/f}$, $\Omega \equiv diag\{1,1,-1,1,1\}$ and
$\tilde \Sigma \equiv \Sigma_0 \Omega \Sigma^\dagger \Omega \Sigma_0$.
$\epsilon_{ijk}$ and $\epsilon_{xy}$ are antisymmetric tensors where
${ijk} =1,2,3$ and ${xy} = 4,5$.  The Lagrangian ${\cal L}_{down}$
that gives mass to the bottom quark will be not be used in our
calculation as we take $\lambda_d \equiv 0$.  The details of this
Lagrangian can be found in Ref.~\cite{belyaev-chen}.  Nevertheless,
we provide Feynman rules for $\lambda_d \neq 0$.

In order to obtain the (exact)
expressions for masses and mixings we will use the vev of the
field $\Sigma$ as given in Eq.~(\ref{sigmavev}), as well as the
vev of the $\xi$ field:
\bea
\langle \xi \rangle_0  \,=\, \left(\begin{array}{ccccc}
1& 0& 0& 0& 0\\
0& \frac{1+c^\prime_v}{2}& \frac{i}{\sqrt{2}}s^\prime_v & 0& 
\frac{c^\prime_v-1}{2}\\
0& \frac{i}{\sqrt{2}}s^\prime_v & c^\prime_v & 0 & 
\frac{i}{\sqrt{2}}s^\prime_v\\
0& 0& 0& 1& 0\\
0& \frac{c^\prime_v-1}{2}& \frac{i}{\sqrt{2}}s^\prime_v & 0 & 
\frac{1+c^\prime_v}{2}\\
\end{array}\right), \label{xivev}
\eea
Where $c^\prime_v \equiv \cos{(\epsilon / \sq2)}$ and
$s^\prime_v \equiv \sin{(\epsilon / \sq2)}$.
The $u^+_{R}$, $U_{R_1}$ and $U_{R_2}$ quark fields are right
handed $SU(2)$ singlets. The upper plus sign in $u^+_R$ denotes
that it is a T-even (T-parity eigenstate) fermion.  With the
other two ($U_{R_1}$ and $U_{R_2}$)
we can define T-even and T-odd linear combinations:
$U^\pm_R \equiv (U_{R_1} \mp U_{R_2})/\sq2$\cite{belyaev-chen}.

On the other hand, the $\Psi$ and $Q$ fields are left handed
$SU(5)$ multiplets defined as:
\bea
\Psi_1 =
\left( \begin{array}{c} 
id_1 \\ -iu_1 \\ 0 \\ 0 \\ 0
\end{array} \right)
\;\;\;\;\;
\Psi_2 =
\left( \begin{array}{c} 
0 \\ 0 \\ 0 \\ id_2 \\ -iu_2
\end{array} \right)
\;\;\;\;\;
Q_1 =
\left( \begin{array}{c} 
id_1 \\ -iu_1 \\ U_{L_1} \\ 0 \\ 0
\end{array} \right)
\;\;\;\;\;
Q_2 =
\left( \begin{array}{c} 
0 \\ 0 \\ U_{L_2} \\ id_2 \\ -iu_2
\end{array} \right)
\eea
From these we define the T-parity eigenstates
$u^\pm_L \equiv (u_{1} \mp u_{2})/\sq2$, 
$d^\pm_L \equiv (d_{1} \mp d_{2})/\sq2$, and
$U^\pm_L \equiv (U_{L_1} \mp U_{L_2})/\sq2$.

The $\Psi_c$ multiplet is composed of 5 right handed T-odd
quark fields: 
\bea
\Psi_c \equiv (\; ib^-\, ,\, -i a^-\, ,\, \chi^- 
\, ,\, i n^- \, ,\, -i p^- \; )^T \label{pax}
\eea
It turns out that two linear combinations of these become the
right handed mass eigenstates of the mirror top and mirror
bottom quarks.
The other three linear combinations are extra T-odd fermions
that are assumed to have very large Dirac masses, so that
they decouple from the main theory\cite{low,hubisz-meade}.

Below we write down the mass eigenstates that arise from
the Lagrangian in Eq.~(\ref{yukawas}) for the top and its
heavy partner:
\bea
\left( \begin{array}{c} 
t^+_{L(R)} \\ T^+_{L(R)} 
\end{array} \right)
=
\left( \begin{array}{cc} 
c_{L(R)} & -s_{L(R)} \\ c_{L(R)} & s_{L(R)}  
\end{array} \right)
\left( \begin{array}{c} 
u^+_{L(R)} \\ U^+_{L(R)} 
\end{array} \right) \; ,
\eea
where $c_{L(R)}=\cos{\theta_{L(R)}}$,
$s_{L(R)}=\sin{\theta_{L(R)}}$, and $\theta_{L(R)}$ is
the mixing angle of the left (right) top and heavy Top quarks.

The mixing angles must satisfy two equations, which we write
in terms of $\tan{\theta_{L(R)}}$:
\bea
r (1+c_v) + \sq2 \, r s_v \tan{\theta_{L}} - 2
\tan{\theta_{R}} &=& 0   \nonumber \\
\sq2 \, r s_v \tan{\theta_{R}} - 2 \tan{\theta_{L}} - 
r(1+c_v) \tan{\theta_{L}} \tan{\theta_{R}} &=& 0 \; ,
\eea
with $r\equiv \lambda_1/\lambda_2$.

The solutions of these equations are:
\bea
\tan{\theta_{L}} &=& \sqrt{a^2_m +1} - a_m \; ,
\nonumber \\
\tan{\theta_{R}} &=& 
\frac{r}{2} \left(1+c_v + \sq2 s_v \tan{\theta_{L}} \right)\; ,
\nonumber \\
a_m &\equiv& \frac{1+c_v-3s^2_v/2+2/r^2}{\sq2 s_v (1+c_v)}
\; .\nonumber
\eea

The masses of the top and its heavy partner are:
\bea
m_{t^+} &=& \frac{1}{2} f\lambda_2 c_L c_R \left[ 
2 \tan{\theta_{L}} \tan{\theta_{R}} + 
r (\sq2 s_v - (1+c_v) \tan{\theta_{L}} ) \right]
\nonumber \\
m_{T^+} &=& \frac{1}{2} f\lambda_2 c_L c_R \left[
2+ r \tan{\theta_{R}} (1+c_v + \sq2 s_v \tan{\theta_{L}} )  
\right] \nonumber
\eea

Expanding in powers of $\epsilon$:
\bea 
\tan{\theta_{L}} &\simeq& \frac{r^2}{1+r^2} \; \epsilon \; +\;
\frac{r^4+2r^2-5}{6 (1+r^2)^3} \; r^2\epsilon^3 \; ,
\nonumber \\
\tan{\theta_{R}} &\simeq& r \; +\; \frac{r^2-1}{1+r^2} \;\frac{r}{2}
\epsilon^2 \; .\nonumber \\
m_{t^+} &\simeq& \frac{f\lambda_1}{\sqrt{1+r^2}}
\left(  \epsilon - 
\frac{2+r^2+2r^4}{6(1+r^2)^2} \epsilon^3 \right)
\nonumber\\
m_{T^+} &\simeq& f\lambda_2 \sqrt{1+r^2}
\left( 1 - \frac{r^2}{2 (1+r^2)^2} \epsilon^2 \right)
\nonumber\\
\cos{\theta_{L}} &\equiv& c_L \; \simeq \;
1-\frac{r^4}{2(1+r^2)^2} \; \epsilon^2 \;.
\label{clexpansion}
\eea
Our formulas for the mixing angles and masses are in agreement with
those of Ref.~\cite{tobe-matsumoto}, where $\alpha \equiv \theta_R$
and $\beta \equiv \theta_L$.

The T-odd top and heavy Top quarks are defined as:
\bea
t^-_L \; =\; u^-_L \; ,\;\;\; && 
\;\;\;\;  T^-_L \; =\; U^-_L \; ,
\nonumber \\
t^-_R \; =\; p^{\prime -}_R  \; ,\;\;\;
&& \;\;\;\; T^-_R \; =\; U^-_R \; .
\eea
Where the $p^{\prime -}$ field comes from the redefinition
of the right handed T-odd fields of Eq.~(\ref{pax}):
\bea
\left( \begin{array}{c} 
p^{\prime -}_R \\ a^{\prime -}_R \\ \chi^{\prime -}_R  
\end{array} \right)
=
\left( \begin{array}{ccc}
\frac{1+c^\prime_v}{2} & \frac{c^\prime_v-1}{2} &
\frac{-1}{\sqrt{2}}s^\prime_v \\
\frac{c^\prime_v-1}{2} & \frac{1+c^\prime_v}{2} &
\frac{-1}{\sqrt{2}}s^\prime_v \\
\frac{1}{\sqrt{2}}s^\prime_v & 
\frac{1}{\sqrt{2}}s^\prime_v & c^\prime_v
\end{array} \right)
\left( \begin{array}{c} 
p^- \\ a^- \\ \chi^- 
\end{array} \right) \; .
\eea
Notice that the mass of $t^-$ comes from the mirror fermion
Lagrangian, whereas the mass of $T^-$ comes from the top quark
Lagrangian (see Eq.~(\ref{yukawas}) ):
\bea
m_{t^-} &=& \sq2 \, f \, \kappa \nonumber \\
m_{T^-} &=& f \, \lambda_2 \; .\nonumber 
\eea

For the calculations in this work we will set $\kappa =1$ so
that the masses of mirror fermions are just $\sq2 f$.
The presence of the LHT mirror fermions is vital for the good
high energy behaviour of the model, in particular they play
an essential role in the scattering process
$u\bar u\to W^+_H W^-_H$ \cite{belyaev-chen}.  Our choice of
$\kappa =1$ and the corresponding values of the T-odd fermion
mass respects the unitarity bounds of this process, as well
as the limits coming from the contributions to the four fermion
contact interaction $e^+e^- \to q\bar q$ \cite{hubisz-meade}.

For completeness, let us write down the masses of the T-even
and T-odd bottom quarks.  The mass of $b^+$ is given by the
down type Yukawa Lagrangian given in Eq.~(\ref{yukawas}).
(See Ref.~\cite{belyaev-chen}):
\bea
m_{b^+} &=& \frac{f}{\sq2} \, \lambda_d \, s_v \, c_v^{-1/4} 
\nonumber \\
m_{b^-} &=& \sq2 \, f \, \kappa \; .\nonumber
\eea
Notice that the formulas we have obtained are exact (to all
orders in the $\epsilon$ expansion); in particular, the mirror
fermion masses are equal for $t^-$ and $b^-$ quarks.  We remind
the reader that in our calculation we take the mass of $b^+$ as
zero ($\lambda_d \equiv 0$).  Feynman rules with the mass
eigenstates can be found in Appendix B.

\section{The $\bar t bW^+$ coupling in the LHT model.}

Let us define the effective $\bar t bW^+$ coupling as follows:
\bea
\mathcal{L}_{\mathrm tbW} &=&  \frac{g}{\sqrt 2}\, W^-_\mu \, 
\bar b \, \gamma^\mu  \left( f_{1L} P_L + f_{1R} P_R  \right)\, t
\nonumber \\
&-& \frac{g}{\sqrt 2 M_W} \,
\partial_\nu W^-_\mu \, \bar b \, \sigma^{\mu\nu}
\left( f_{2L} P_L + f_{2R} P_R \right) \, t \;\;\;+\; h.c.\, ,
\label{tbwvertex} 
\eea
where we have used the mass scale $m_W$ that is also used
in the literature~\cite{chen-tpol,kane,aguila}.

\begin{figure}
\includegraphics[width=10.5cm,height=8.5cm]{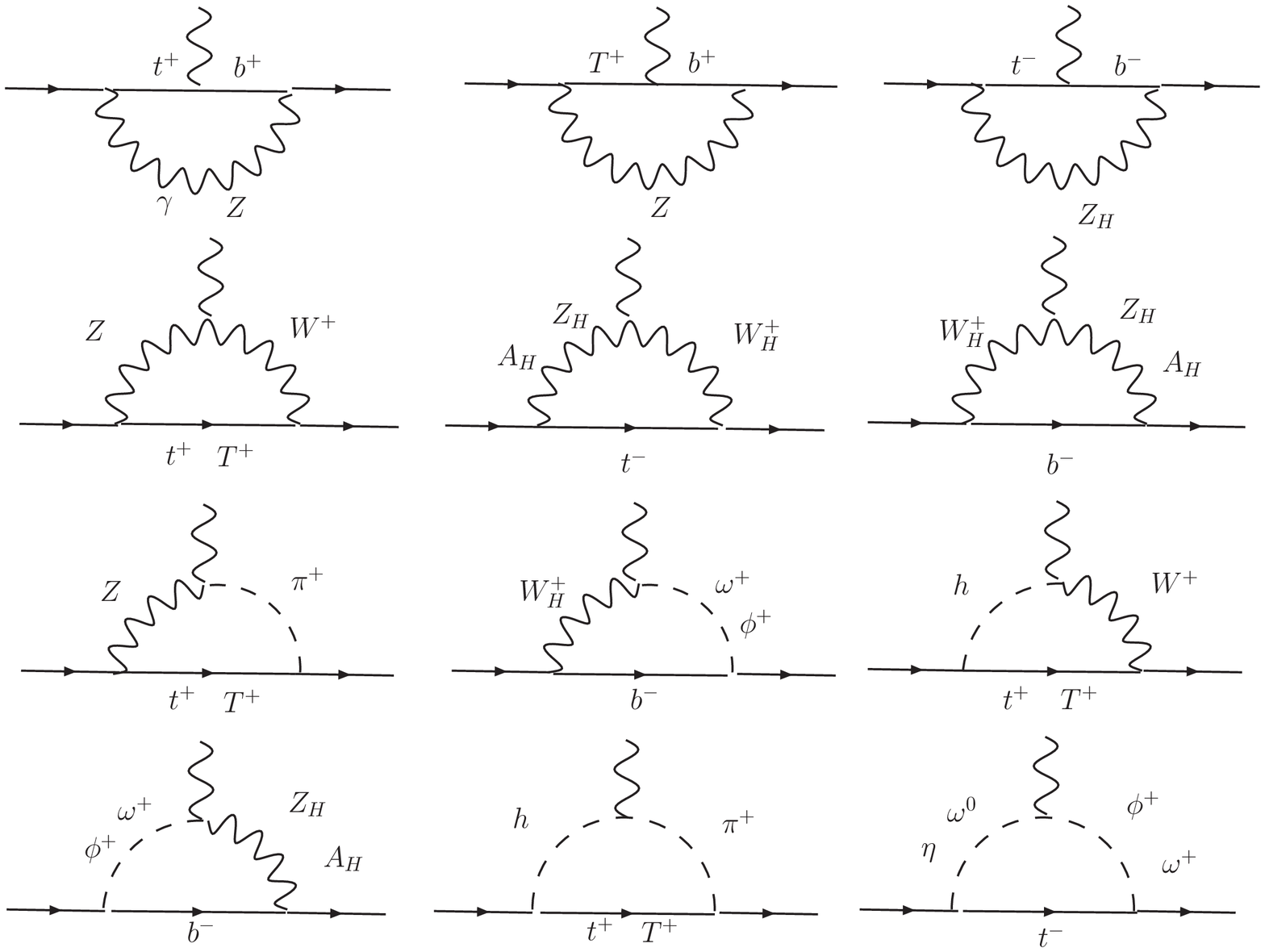}
\caption{Some Feynman diagrams that give rise to the
$f_{2R}$ coupling of the $tbW$ vertex in the LHT model.}
\label{diagrams}
\end{figure}

In the SM the values of the form factors at tree level are
$f_{1L} = V_{tb}\simeq 1$, $f_{1R} = f_{2L} = f_{2R} = 0$.
Radiative corrections to the factors $f_{1R}$ and $f_{2L}$ 
must be zero if we neglect the mass of the bottom quark.
We take $m_b\equiv 0$ in this work, so we set
$f_{1R}=f_{2L}\equiv 0$ for this study. 
These couplings can be probed by
studying the top decay $t\to bW^+$ and the single top
production processes\cite{chen-tpol,bernreuther}.
The dimension five coupling $f_{2R}$ is different from zero at
one loop: we obtain $f^{\rm SM}_{2R} = 0.00201$ ($0.00214$) for
$m_H = 120$ ($150$) GeV.  This value seems to be too small to be
probed at the LHC\cite{aguilartbw}.
In fact, the dominant radiative corrections for the top width
or single top production comes from QCD\cite{bernreuther}.
We would like to know if the coeficient $f_{2R}$ predicted by
the LHT model could be large enough to be measured at the LHC.

In the LHT model, the coefficient $f_{1L}$ is modified at
tree level by the $tT$ mixing angle $\theta_L$ ($f_{1L}=c_L$).
The tree level $tbW$ vertex is reduced by the factor $c_L$ and
this translates into a lower production at the LHC \cite{cao-li}.

We have performed the one loop contribution to $f_{2R}$ in
the LHT model.   We have worked in the Feynman-t'Hooft gauge,
where there are a total of 47 diagrams to compute if we take
the bottom quark mass as zero. Some of the diagrams are shown
in Figure~(\ref{diagrams}).  (We provide the Feynman rules
necessary for such a calculation in Appendix B.) 
Notice that the Goldstone bosons in the original
Lagrangian have to be diagonalized and normalized.
We have done all this exactly (at all orders in powers of
$\epsilon$) in Appendix A.  The exponential expansion in
the Lagrangians of Eqs. (\ref{kineticlagrangian}) and
(\ref{yukawas}) generate vertices of dimension 4 and higher
that contribute at one loop to $f_{2R}$.  As it turns out,
the contribution from the dimension 4 terms to the $tbW$
vertex are finite, whereas
the contribution from the higher dimension terms is divergent.
This is no surprise because the LHT model is a non-renormalizable
effective low energy model with a cut-off scale
($\Lambda \sim 4 \pi f$).   In principle, all the operators
that are consistent with the symmetries of the LHT model should
be considered\cite{hubisz-precision}.  In our study,
we disregard effects from higher dimension terms and keep only
the contribution from the dimension 4 couplings that render a
finite result\cite{foot1}.

Concerning the specific numerical values used for the parameters of
the model, we have chosen values from the allowed region of $f$ vs
$\lambda_2$ that is shown in Fig. 1 of Ref.~\cite{tobe-matsumoto}.
The mass of the top quark is taken $m_t = 173$ GeV, and this sets
the value of $\lambda_1 \simeq 1$ (with a very small dependence
on the value of $f$).  The masses of the $t^-$ and $b^-$
mirror quarks is taken as $\sq2 f$.  The masses of the physical
T-odd scalars $\phi^{\prime \pm}$, $\phi^{\prime 0}$, and
$\phi^{\prime 0}_P$ are taken as $\sq2 m_H f/v$
($=0.69 f$ for $m_H=120$ GeV) as it is done in the
literature\cite{han-logan, hubisz-meade}. As it is well known,
in the Feynman-t'Hooft gauge
the masses of Goldstone bosons ($\pi^{\prime \pm}$, $\pi^{\prime 0}$,
$\omega^{\prime \pm}$, $\omega^{\prime 0}$ and $\eta^{\prime}$)
are equal to the masses of their corresponding
gauge bosons which are given in Eq.~(\ref{bosonmasses}).
We have chosen a range of the scale $f$ between $550$ GeV and
$1550$ GeV.  Smaller values of $f$ are prohibited by the
low energy data\cite{hubisz-precision}.  Higher values are
allowed but not interesting as the value of $f_{2R}$ remains
essentially constant above the $1.5$ TeV scale.

\begin{figure}[here]
\includegraphics[width=9.0cm,height=7.5cm]{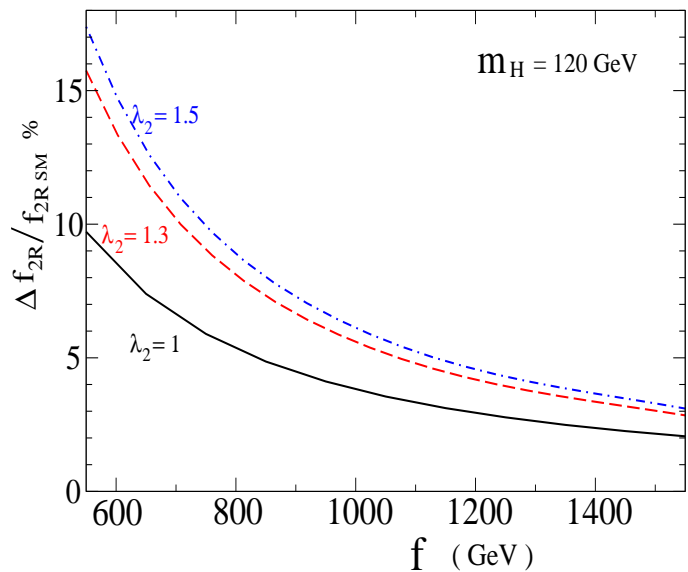}
\caption{The $f_{2R}$ variation for $m_H=120$ GeV.}
\label{f2r120}
\end{figure}

\begin{figure}[here]
\includegraphics[width=9.0cm,height=7.5cm]{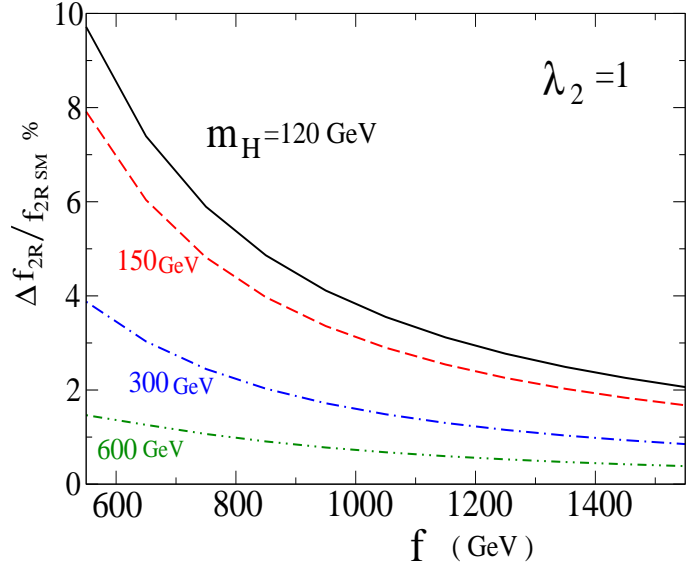}
\caption{The $f_{2R}$ variation for several values of $m_H$.}
\label{f2r150}
\end{figure}

The variation of $f_{2R}$ from $f^{\rm SM}_{2R}$ as predicted
by the LHT model turns out to be of the order expected by a one
loop correction. In fact, $\Delta f_{2R}/f^{\rm SM}_{2R}$ is
under $20 \%$ for the allowed values of the scale $f$ and
the Yukawa coefficient $\lambda_2$.  We show the variation of
$f_{2R}$ as a function of the scale $f$ in Fig.~(\ref{f2r120}).
We also show in Fig.~(\ref{f2r150}) how the variation in
$f_{2R}$ diminishes with higher values of the Higgs mass.
In contrast to the SM the LHT model allows for higher values
of $m_H$ and is still compatible with the electroweak precision
data\cite{hubisz-precision}, however we have found that for
bigger $m_H$ the deviation in $f_{2R}$ gets smaller and thus
less interesting.  (Observe in Fig.~\ref{diagrams} that the
Higgs field also appears in the contributions from the LHT
heavy states.)  From now on we will assume a fixed value
$m_H=120$ GeV.

It is possible to obtain the variation in the top width, the
W-helicity in the $t\to bW^+$ decay, as well as the s and t
channels of the single top production processes once we have
the effective $tbW$ coupling.  A general analysis of this
coupling and the observables mentioned has been done
in Ref.~\cite{chen-tpol}.  Let us apply this approach to the
effective $tbW$ coupling as predicted by the LHT model.  The
total $t\to b W^+$ decay width of the top quark can be
written as a sum of the contributions from each of the three
polarizations of the $W^+$ boson:
\bea
\Gamma_t &=& \Gamma_0 + \Gamma_- +\Gamma_+ \nonumber \\
&=& \frac{g^2 m_t}{64\pi} \; \frac{(a_t^2-1)^2}{a_t^4}
\left( a_t^2 L_0^2 + 2T_m^2 + 2T_p^2 \right)\, ,
\nonumber \\
L_0^2 &=& (f_{1L} + f_{2R}/a_t)^2 + (f_{1R} + f_{2L}/a_t)^2 \, ,
\nonumber \\
T_m^2 &=& (f_{1L} + a_t f_{2R})^2 \, ,
\nonumber \\
T_p^2 &=& (f_{1R} + a_t f_{2L})^2\, ,
\label{width} \\
a_t &=& \frac{m_t}{m_W}\, . \nonumber
\eea
From this expression we define the W-helicity ratios
$$f_0 \equiv \frac{\Gamma_0}{\Gamma_t} \; ,\;\;\;\;\;
f_- \equiv \frac{\Gamma_-}{\Gamma_t} \; ,\;\;\; {\rm and}
\;\;\; f_+ \equiv \frac{\Gamma_+}{\Gamma_t}\; .$$
Notice that the $T_p$ coefficient is zero for $m_b=0$. However,
we are including it here for the sake of completeness.  For
$f_+ = 0$ we have that $f_0 + f_-$ must be equal to one.
Therefore, it is only necessary to study one of them.  In this
work we show the deviation in $f_-$ predicted by the LHT model.

It is convenient to define the following effective terms:
\bea
x_0 &=& L_0^2 - 1 \, , \;\;\;\;\;\;\;
x_5 \; =\; a_t^2 (f_{2R} + f_{2L})^2 \, ,
\nonumber \\
x_m &=& T_m^2 - 1 \, , \;\;\;\;\;\;
x_p \; =\; T_p^2  \, .
\label{xterms}
\eea

Then, the W-helicity ratios and the single top production
cross section for the s and t channels are given by
\bea
f_0 &=& \frac{a^2_t (1+x_0)}{a^2_t(1+x_0)+2(1+x_m+x_p)}\, ,
\nonumber \\
f_+ &=& \frac{2x_p}{a^2_t(1+x_0)+2(1+x_m+x_p)}\, ,
\nonumber \\
f_- &=& \frac{2(1+x_m)}{a^2_t(1+x_0)+2(1+x_m+x_p)}\, ,
\nonumber \\
\sigma_t &=& \sigma^{\rm SM}_t\, +\,
 a_0 x_0 +a_m x_m + a_p x_p + a_5 x_5\, ,
\nonumber \\
\sigma_s &=& \sigma^{\rm SM}_s\, +\,
b_0 x_0 +b_m x_m + b_p x_p + b_5 x_5\, .
\label{sigmas}
\eea
The numerical values of the $a_j$ and $b_j$ coefficients
are given in Ref.~\cite{chen-tpol} for a mass of the top
quark $m_t = 178$ GeV.  In Table~(\ref{abcoefs}) we show their
values for $m_t = 173$ GeV.  We have used the CTEQ6L1 parton
distribution function
when integrating over the parton luminosities\cite{pdf}.

\begin{figure}[here]
\includegraphics[width=9.0cm,height=7.5cm]{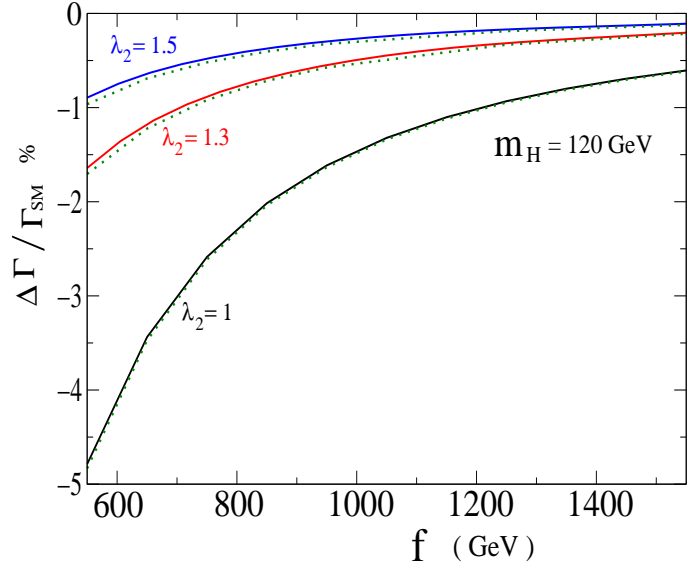}
\caption{The variation in the top width. Solid lines contain
the contribution from both the $f_{1L}$ and $f_{2R}$ couplings.
Dotted lines are for $f_{2R}=0$. }
\label{dwidth}
\end{figure}

\begin{figure}[here]
\includegraphics[width=9.0cm,height=7.5cm]{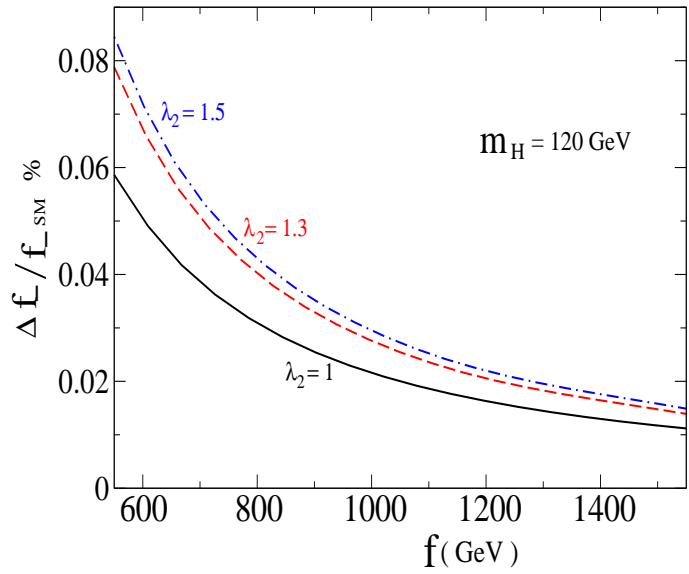}
\caption{The variation in the W-boson helicity ratio
$f_-$. It would require a deviation of $f_{2R}$ much
bigger than $20\%$ in order to have a significant change
in $f_-$.}
\label{dfm}
\end{figure}

\begin{table}[ht]
\begin{tabular}{|c||c|c|c|c|c|}
\hline t-channel: & $a_0$ & $a_m$ & $a_p$ & $a_5$ & 
$\sigma^{\rm SM}_t$
\tabularnewline \hline \hline
Tevatron & 0.995 & -0.089 & -0.181 & 0.336 &
0.906
\tabularnewline \hline 
LHC ($t$) & 174.2 & -22.2 & -38.1 & 78.2 &
151.99
\tabularnewline \hline 
LHC ($\bar t$) & 111.9 & -23.4 & -14.6 & 48.7 &
88.46 
\tabularnewline \hline \hline
s-channel: & $b_0$ & $b_m$ & $b_p$ & $b_5$ &
$\sigma^{\rm SM}_s$
\tabularnewline \hline \hline
Tevatron & -0.094 & 0.040 & 0.040 & 0.263 &
0.306
\tabularnewline \hline 
LHC ($t$) & -1.58 & 6.30 & 6.30 & 7.02 &
4.716
\tabularnewline \hline 
LHC ($\bar t$) & -0.944 & 3.83 & 3.83 & 3.76 &
2.884
\tabularnewline \hline
\end{tabular}
\caption{The single top production cross section coefficients 
of Eq.~(\ref{sigmas}) for $m_t = 173$ GeV.  Last column is
the Born level production cross section in the SM.
All in units of pb.
\label{abcoefs}}
\end{table}

As mentioned above, the LHT model predicts a tree level
reduction of the $tbW$ coupling.  Therefore, an important
feature of this model is that at tree level all single
top production modes
as well as the total decay width show the same proportional
deviation from the SM prediction\cite{cao-li}.  In our study,
we want to consider the additional effect from the dimension-5
$f_{2R}$ coupling that arises at the one loop level in LHT.

The SM born level prediction of the $t\to bW^+$ width of the
top quark is $\Gamma (t\to bW^+)=1.5$ GeV for $m_t=173$ GeV.
There is a $10\%$ decrease when QCD and electroweak corrections
as well as non-zero $m_b$ and finite W-boson width effects
are considered\cite{korner, bernreuther}.
In Fig.~(\ref{dwidth}) we show
the deviation in the total width of the top quark coming from
the LHT model.   The solid lines in Fig.~(\ref{dwidth}) 
give the reduction
in $\Gamma_t$ as a funtion of the scale $f$ and three different
values of $\lambda_2$.  For these lines both the effective
$f_{1L}$ and $f_{2R}$ couplings are considered.  Nonetheless,
we also show in dotted lines the same curves obtained when
only the $f_{1L}$ coupling is considered.  Dotted and solid
lines almost overlap: as expected, the change in $\Gamma_t$
is driven mainly by the tree level mixing with the heavy top.  
We conclude that the
small changes in $f_{2R}$ cannot be seen by measuring $\Gamma_t$.
Also, notice that the ($\%$) reduction  in Fig.~(\ref{dwidth}) is
entirely due to the cosine of the $tT$ mixing angle $c_L$, which
according to formula (\ref{clexpansion}) tends to 1 when
either $\epsilon \to 0$ or $r=\lambda_1/\lambda_2 \to 0$.

As for the W-boson helicity ratios $f_0$ and $f_-$, in
principle, these observables are more sensitive to the
$f_{2R}$ coupling.  Notice that $L_0$ and $T_m$ in
equation~(\ref{width}) get exactly the same correction if only
the $f_{1L}$ is modified (we have set $f_{1R} = f_{2L} =0$).
This means that the ratios $f_0$ and
$f_-$ do not change at all from their SM values when we
consider only the tree level $tbW$ coupling of the LHT model.
However, when we consider the change in the $f_{2R}$ coupling
we do observe a deviation that (unfortunately) turns out to
be very small (of order less than $0.1\%$) as it is shown in
Fig.~(\ref{dfm}).  We conclude that the W-boson helicity ratios
$f_0$ and $f_-$ require a substantial deviation in the dimension
five coupling $f_{2R}$ in order to show significant changes
from their SM values.

\begin{figure}
\includegraphics[width=9.0cm,height=7.5cm]{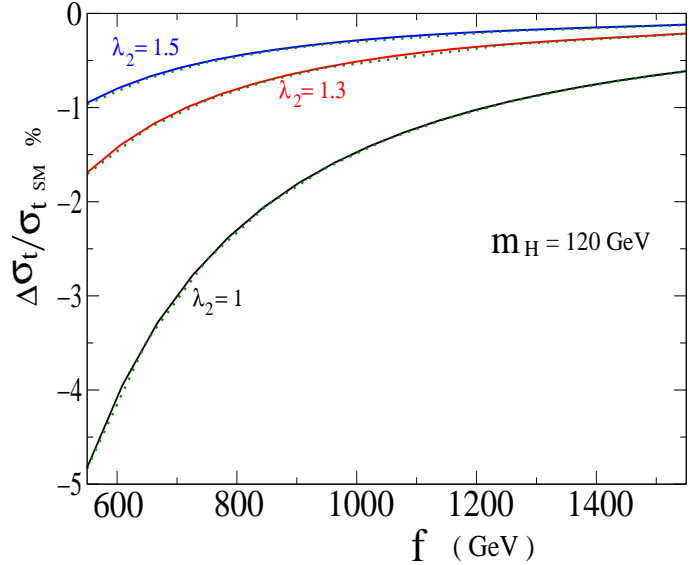}
\caption{The variation in the $\sigma_t$ cross section.
Solid lines contain the contribution from both the $f_{1L}$ 
and $f_{2R}$ couplings. Dotted lines are for $f_{2R}=0$.}
\label{dsigmat}
\end{figure}

\begin{figure}
\includegraphics[width=9.0cm,height=7.5cm]{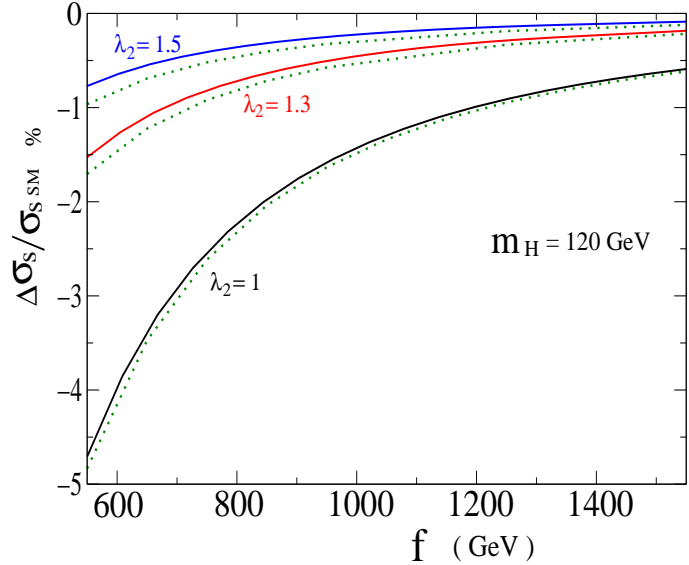}
\caption{The variation in the $\sigma_s$ cross section.
Solid lines contain the contribution from both the $f_{1L}$ 
and $f_{2R}$ couplings. Dotted lines are for $f_{2R}=0$.}
\label{dsigmas}
\end{figure}

As mentioned above, the effective couplings $f_{1L}$ and $f_{2R}$
can also be probed with single top production.  In comparison with
the top decay width $\Gamma_t$ and the W-helicity ratio $f_-$, the
cross section could be more sensitive to the $f_{2R}$ coupling.
We show the deviation in the t-channel cross section
in Fig.~(\ref{dsigmat}).  Notice that the change when we go from
considering the deviation in $f_{1L}$ only (dotted lines), to
considering  both the deviations in $f_{1L}$ and $f_{2R}$ 
(solid lines) is hardly visible.  
This change is slightly more pronounced for the s-channel cross
section as shown in Fig.~(\ref{dsigmas}).   We can observe from the
values of the effective coefficients in Table~(\ref{abcoefs}) that
for this channel the $f_{2R}$ coupling has a somewhat bigger effect
through the $x_5$ and $x_0$ terms defined in equation~(\ref{xterms}). 
For instance, for a scale $f=550$ GeV and $\lambda_2 =1.5$ the tree
level $f_{1L}$ reduction expected in the LHT model brings about 
a $1.0\%$ reduction in $\sigma_s$ (dotted line), 
whereas the combined $f_{1L}$ and $f_{2R}$ deviations bring a 
smaller $0.8\%$ reduction in $\sigma_s$ (solid line).
The reason for this can be seen in Fig.~(\ref{f2r120}).  The LHT
contribution, on the one hand, decreases the value of $f_{1L}$,
and on the other hand increases the (positive) value of $f_{2R}$
with respect to the SM.  We thus have a small compensation
in the value of the $x$ terms in Eq.~(\ref{xterms}).

\section{Conclusions}

Besides the SM electroweak parameters, the LHT model adds
three more free parameters: $\kappa$, $\lambda_2$ and the
scale $f$ (which is associated with an estimated cut-off
$\Lambda \sim 4\pi f$).  We have chosen a value of the mirror
fermion yukawa $\kappa =1$ that for our range of $f$ gives T-odd
fermion masses that are consistent with bounds from four
fermion contact interactions $e^+e^-q\bar q$ and from
unitarity in $u\bar u \to W^+_H W^-_H$ scattering processes.
Our study concentrates on the two other parameters: $\lambda_2$
(which drives $m_{T^+}$) and the scale $f$.   As for the values
of $\lambda_2$ and $f$ we have chosen $\lambda_2 =1$, $1.3$ and
$1.5$ and $550 \leq f \leq 1550$ GeV as suggested by
Ref.~\cite{tobe-matsumoto}.

Because of the mixing between the top quark and its heavy
partner $T^+$,
the LHT model predicts a tree level reduction of the dim 4 
$f_{1L}$ coupling that implies an expected proportional
reduction in the top width and single top production.
Changes in $f_{1L}$ by themselves do not vary the
predicted W helicity fractions in the $t\to bW$ decay.  However,
the contribution from the $f_{2R}$ coupling could modify these
fractions. The dim 5 weak dipole coupling $f_{2R}$ arises at the
one loop level in the SM and in the LHT model.  It is somewhat
increased in size by the LHT model.  However, the contribution
to $f_m$ (and $f_0$) is negligible. The increase in $f_{2R}$
predicted by the LHT tends to counteract the effects of the
reduction in $f_{1L}$.  In any case, the effects from $f_{2R}$
are very small. We have found that the tree level analysis of
the top's width and the single top cross section remains valid
for the LHT model.  Top quark observables like total width,
and single top production are sensitive to the mixing with
the heavy $T^+$ partner.   In particular, the LHC may probe
deviations of the $f_{1L}$ $tbW$ coupling down to a few
percent\cite{aguilartbw} and this would imply an indirect probe
of the scale $f$ (and the yukawa $\lambda_2$)  of the LHT model
( see Eq.~(\ref{clexpansion}) ).   Of course, there are direct
tests of the new heavy states at the LHC that will give more
precise determination of these parameters.  A recent study
(see Ref.~\cite{tobe-matsumoto}) has shown that signal events
from $T^+$ and $T^-$ production can be distinguished from SM
backgrounds so that the mass and mixings of the top partners
can be obtained with relatively good accuracies. Furthermore,
other studies have shown that, since the mass of the T-odd
fermions cannot be too heavy to be consistent with low energy
data, they can be produced at high enough rates at the
LHC\cite{belyaev-chen}.


\vspace*{4cm}

\noindent
{\bf Acknowledgments}~~~
We thank C.-P. Yuan, Chuan-Ren Chen, M.A. P\'erez and R. Martinez
for useful discussions.  We thank Conacyt for support.

\newpage

\appendix

\section{The Goldstone boson sector in the 't Hooft-Feynman gauge.}

In the LHT model, the charged fields $\omega^{\pm}$ and $\phi^\pm$
as well as the neutral fields $\omega^0$, $\eta$ and $\phi^0_P$, 
mix at order $(\frac{v}{f})^2$. It is a linear combination of these
that is eaten by the heavy gauge bosons when the extended gauge
group is broken down to $SU(2)_L\times U(1)_Y$.  On the other hand,
the $\pi$ fields are T-parity even and do not mix with the other
scalars. They are absorbed by the standard model $W/Z$ bosons as usual.  
The fields $h$, $\phi^0$, $\phi^0_P$ and $\phi^\pm$ remain in the
spectrum (after diagonalization).
The basis of the kinetic Lagrangian of 
Eq.~(\ref{kineticlagrangian}) is an exponential matrix that is
usually computed up to the first few leading terms.
However, it is possible to obtain the exact expressions for
the kinetic ($\partial^\mu \phi \partial_\mu \phi$) scalar,
the scalar-boson mixing ($W_\mu^+\partial^\mu \pi^-$) and 
boson mass terms. (It is from the latter that the boson masses
of Eq.~(\ref{bosonmasses}) were obtained.)

In obtaining the following formulas it is convenient to notice that
the vev value of the field matrix $\Pi$ Eq.~(\ref{14pions}) is
proportional to a matrix $M_0$:
\bea
M_0 \,= \left(\begin{array}{ccccc}
~0~ & ~0~ & ~0~ & ~0~ & ~0~ \\
0 & 0 & 1 & 0 & 0 \\
0 & 1 & 0 & 0 & 1 \\
0 & 0 & 0 & 0 & 0 \\
0 & 0 & 1 & 0 & 0 
\end{array}\right), \label{mzero}
\eea
for which is easy to prove that $M_0^{2n+1}=2^n M_0$ and
$M_0^{2(n+1)}=2^n M_0^2$ with $n=0,1,2,3,..$.  With these
identities it can be shown that the vev of $\Sigma$
is\cite{hubisz-precision}:
\bea
\langle \Sigma \rangle_0 \,=\, \left(\begin{array}{ccccc}
0& 0& 0& 1& 0\\
0& -\frac{1-c_v}{2}& \frac{i}{\sqrt{2}}s_v& 0& \frac{1+c_v}{2}\\
0& \frac{i}{\sqrt{2}}s_v& c_v& 0& \frac{i}{\sqrt{2}}s_v\\
1& 0& 0& 0& 0\\
0& \frac{1+c_v}{2}& \frac{i}{\sqrt{2}}s_v& 0& -\frac{1-c_v}{2}\\
\end{array}\right), \label{sigmavev}
\eea
where $s_v=\sin{\sqrt{2}\epsilon}$ and $\epsilon \equiv v/f$ as
defined in Eq.~\ref{epsilon}.  This expression can be used
in the kinetic Lagrangian Eq.~(\ref{kineticlagrangian}) to obtain
the (exact) mixing and mass terms of the LHT model.
The diagonalization and normalization of Goldstone boson fields
has been discussed at order $\epsilon^2$ for the charged sector
in Ref.~\cite{hubisz-precision} and for the neutral sector in
Ref.~\cite{buras2}.  Below, we will make the
same analysis for charged and neutral bosons exactly (at all
orders in $\epsilon$).

\subsection{The charged $W^{\pm}$ bosons.}

Let us write down the part of the kinetic Lagrangian
Eq.~(\ref{kineticlagrangian}) that involves the charged
bosons of the LHT model.  It is convenient to put it in
a matricial form:

\bea
{\cal L}_{\rm kin} &=& \partial^\mu \phi^{++} \partial_\mu \phi^{--}
\; +\; 2 \tilde \kappa \; \partial^\mu \pi^+
\partial_\mu \pi^- \; +\; fg\frac{1-c_v}{\sq2\; y}
(W^+_\mu \partial^\mu \pi^- \; +\;W^-_\mu \partial^\mu \pi^+ )
\nonumber \\
&+& ( \; \partial_\mu \omega^- \;\;\;\; \partial_\mu \phi^- \;)
\; B\; 
\left(\begin{array}{c}
\partial^\mu \omega^+   \\
\partial^\mu \phi^+
\end{array}\right)
\; +\; fg W^+_{H\mu}
( \;\frac{1+\kappa_0}{2} \;\;\;\; i\frac{1-\kappa_0}{2} \;)\; 
\left(\begin{array}{c}
\partial^\mu \omega^-   \\
\partial^\mu \phi^-
\end{array}\right)
\nonumber \\
&+& fg W^-_{H\mu}
( \;\frac{1+\kappa_0}{2} \;\;\;\; -i\frac{1-\kappa_0}{2} \;)\; 
\left(\begin{array}{c}
\partial^\mu \omega^+   \\
\partial^\mu \phi^+
\end{array}\right)
\eea
with
\bea
B &=&
\left(\begin{array}{cc}
\frac{1}{2}+\tilde \kappa & 
-i(\frac{1}{2}-\tilde \kappa)  \\
i(\frac{1}{2}-\tilde \kappa) & 
\frac{1}{2}+\tilde \kappa  
\end{array}\right)
\nonumber \\
\kappa_0 &=& \frac{s_v}{\sqrt{2}\epsilon} \;=\; 
1-\frac{2\epsilon^2}{3!}+
\frac{4 \epsilon^4}{5!} -... \nonumber \\
\tilde \kappa &=& \frac{1-c_v}{2\epsilon^2} \;=\;
\frac{1}{2!}-\frac{2\epsilon^2}{4!}+\frac{4\epsilon^4}{6!}-
 ...\label{k0}
\eea
We then redefine the T-even charged scalar $\pi^\pm$
as well as the T-odd $\omega^\pm$ and $\phi^\pm$ to
diagonalize the Lagrangian.
The new T-even $\pi^{\prime \pm}$ field is given by: 
\bea
\pi^{\pm} &\equiv& \; \frac{\pm \, i}{\sqrt{2\tilde \kappa}} 
\; \pi^{\prime \pm} \label{piprime}
\eea
 An extra phase $i$ multiplies the $\pi^{\prime \pm}$ field so
that the $W^\pm \pi^{\prime \pm}$ mixing and the $SU(2)$
gauge-fixing terms become identical to the usual SM
expressions\cite{hollik}.
The other two charged scalars are redefined as:
\bea
\left(\begin{array}{c}
\omega^+   \\
\phi^+
\end{array}\right) \; \equiv \; \; T \; 
\left(\begin{array}{c}
i\omega^{\prime +}   \\
\phi^{\prime +}
\end{array}\right)
\;\;\;\;\;\;
T \; =\; \frac{\hat m_{W_H}}{2\tilde \kappa + \kappa_0^2}
\left(\begin{array}{cc}
2\tilde \kappa + \kappa_0 & 
i(1-\kappa_0) \sqrt{2\tilde \kappa}  \\
i(2\tilde \kappa -\kappa_0 ) & 
(1+\kappa_0) \sqrt{2\tilde \kappa}
\end{array}\right)
\eea
where $\hat m_{W_H} = \frac{M_{W_H}}{fg} \; = \frac{\sqrt{3+c_v}}{2}$.
Notice that the new $\omega^{\prime \pm}$ field has an extra phase
factor $\pm i$ that is convenient to use so that the Feynman rules
of this T-odd Goldstone boson resemble the rules of its T-even
counterpart (the $\pi^{\prime \pm}$ boson).  On the other hand,
for the physical heavy T-odd $\phi^{\prime \pm}$ boson we choose
not to insert the phase factor.  The Feynman rules in appendix B
stand for the new $\pi^{\prime \pm}$, $\omega^{\prime \pm}$,
$\phi^{\prime \pm}$, etc. fields, but we have dropped the $\prime$
symbol for simplicity.

In terms of the new (mass eigenstates) fields the Lagrangian becomes:
\bea
{\cal L}_{\rm kin} &=& 
 \partial^\mu \phi^{\prime +} \partial_\mu \phi^{\prime -} 
\nonumber \\
&+& \partial^\mu \pi^{\prime +} \partial_\mu \pi^{\prime -}
 \; +\; i M_{W} 
\left(  W^-_{\mu} \partial^\mu \pi^{\prime +} -
W^+_{\mu} \partial^\mu \pi^{\prime -} \right) \nonumber \\
&+& \partial^\mu \omega^{\prime +}\partial_\mu \omega^{\prime -} 
\; +\; i M_{W_H} \left(  W^-_{H\mu} \partial^\mu \omega^{\prime +}
- W^+_{H\mu} \partial^\mu \omega^{\prime -} \right)
\eea
where the $W^+_{H\mu} \partial^\mu \omega^{\prime -}$ mixing
term is canceled (after integration by parts) when we add
the usual gauge-fixing term:
\begin{equation}
\Delta {\cal L} =
\frac{-1}{\xi_L} \left| \partial^\mu W_{\mu}^\pm -i M_W \xi_L 
\pi^{\prime \pm} \right|^2 \;-\; 
\frac{1}{\xi_H} \left| \partial^\mu W_{H\mu}^\pm -i 
M_{W_H} \xi_H \omega^{\prime \pm} \right|^2.
\end{equation}

To obtain Feynman rules it is convenient to use an expansion
in powers of $\epsilon = v/f$:
\bea
T \; =\; {\bf 1} \;+\;
\left(\begin{array}{cc}
1 & 4i  \\
2i & 1 
\end{array}\right) \;
\frac{\epsilon^2}{24} \;\; +\;
\left(\begin{array}{cc}
-\frac{61}{8} & 13i  \\
\frac{19}{2}i & -1 
\end{array}\right)  \frac{\epsilon^4}{720}
\eea

\subsection{The neutral bosons.}

The neutral boson sector in the Lagrangian of
Eq.~(\ref{kineticlagrangian}) can also be written in the
following matricial form (notice that
$M_Z=\frac{vg}{2c_w} \sqrt{2\tilde \kappa}$):
\bea
{\cal L}_{\rm kin} &=& \frac{1}{2} (\partial^\mu h)^2 \; +\;
\tilde \kappa (\partial^\mu \phi^0)^2 \;+\;
\tilde \kappa (\partial^\mu \pi^0)^2 \; +\; 
\frac{vg}{c_w} \tilde \kappa \; Z^\mu \partial_\mu \pi^0
\nonumber \\
&+& \bar x^T \, A \;\bar x \; +\;  A_H^\mu \;
\bar a^T \bar x \;+\; Z_H^\mu \;\bar b^T \bar x  
\eea
with
\bea
\bar x \; =\; \left(\begin{array}{c}
\partial_\mu \omega^0   \\
\partial_\mu \eta   \\
\partial_\mu \phi^0_P
\end{array}\right)
\;\;\;
\bar a &=& \frac{f g c_H}{8}\left(\begin{array}{c}
8t_w - (4\kappa_1 + \kappa_2) x_\omega \\
(8 t_w -5 \kappa_1 x_\omega )/{\sqrt{5}}   \\
\sq2 \kappa_1 x_\omega
\end{array}\right)
\;\;\;
\bar b \;=\; \frac{f g c_H}{8}\left(\begin{array}{c}
8 - \kappa_1 x_\eta \\
(8 - \kappa_2 x_\eta)/{\sqrt{5}}   \\
-\sq2 \kappa_1 x_\eta
\end{array}\right)
\nonumber \\
A &=& \frac{1}{16}\;
\left(\begin{array}{ccc}
7+\kappa_0^2 & 
\sqrt{5} (1-\kappa_0^2) &
-\sqrt{2} (1-\kappa_0^2) \\
\sqrt{5} (1-\kappa_0^2) & 
3+5\kappa_0^2 &
\sqrt{10} (1-\kappa_0^2)  \\
-\sqrt{2} (1-\kappa_0^2) & 
\sqrt{10} (1-\kappa_0^2)  &
6+2\kappa_0^2
\end{array}\right)
\nonumber 
\eea
and
\bea
\kappa_1 \; =\; 1-\kappa_0 c_v \;\;
\;\;\;\;\;\;\;\;\;\;\;\;\;\;\;\;\;\;\;\;\;\;\; &&
\kappa_2 \; =\; 3+5\kappa_0 c_v  \nonumber \\
x_\omega \; =\; t_w + t_H  
\;\;\;\;\;\;\;\;\;\;\;\;\;\;\;\;\;\;\;\;\;\;\; &&
x_\eta \; =\; 1-t_w t_H 
\nonumber \\
y_\omega \; =\; t_w - 5 t_H 
\;\;\;\;\;\;\;\;\;\;\;\;\;\;\;\;\;\;\;\;\;\;\; &&
y_\eta \; =\; 5 + t_w t_H   \label{xys} 
\eea
Where $t_w \equiv s_w/c_w$ and $t_H$ are Tan($\theta_w$) and
Tan($\theta_H$) respectively ( see Eq.~(\ref{thequation}) ).
To normalize the $\pi^0$ and $\phi^0$ fields we simply
redefine $\pi^{\prime 0} \equiv \sqrt{2\tilde \kappa}\pi^0$
and  $\phi^{\prime 0} \equiv \sqrt{2\tilde \kappa}\phi^0$.
(The Higgs field $h$ needs no redefinition.)
We also redefine the other scalars to properly diagonalize the
Lagrangian:
\bea
\left(\begin{array}{c}
\omega^0   \\
\eta \\
\phi^0_P
\end{array}\right) \; \equiv \;\; T \; 
\left(\begin{array}{c}
\omega^{\prime 0}   \\
\eta^{\prime} \\
\phi^{\prime 0}_P
\end{array}\right)
\;\;\;\;\; {\rm with} \;
T \; =\; \frac{1}{\kappa_0 \sq2 \sqrt{1+3c^2_v}}
\; D_1 \; \left( \;t_{ij} \; \right) \; D_2
\label{tnc}
\eea
where $T$ is conveniently written as a product of 3 matrices:
$D_1$, $(t_{ij})$ and $D_2$.  Two of them are diagonal matrices
defined as $D_1=diag(1,\sqrt{5},\sqrt{2})$, and
$D_2=diag(1/d_\omega,1/d_\eta,1/2)$, with
\bea
 d_\omega &=& \sqrt{8(t^2_w+5t^2_H)-5 s^2_v x^2_\omega} \;,
\; {\rm and} \nonumber \\
d_\eta &=& \sqrt{8(5+t^2_w t^2_H)-5s^2_v x^2_\eta}\; .
\nonumber 
\eea

The matrix elements $t_{ij}$ are as follows:
\bea
t_{11} \;=\; \kappa_0 (5c^2_v x_\omega+2t_w) + c_v y_\omega
\;\;\;\;\; && 
t_{12} \;=\; \kappa_0 (5c^2_v x_\eta-2t_wt_H) - c_v y_\eta
\;\;\;\;\;\;\;\;
t_{13} \;=\; \kappa_1 \nonumber \\
t_{21} \;=\; \kappa_0 (c^2_v x_\omega+2t_H) - c_v y_\omega
\;\;\;\;\;\;\; && 
t_{22} \;=\; \kappa_0 (c^2_v x_\eta+2) + c_v y_\eta
\;\;\;\;\;\;\;\;\;\;\;\;\;\;\;\;\;
t_{23} \;=\; -\kappa_1 \nonumber \\
t_{31} \;=\; -(\kappa_0 - c_v) y_\omega \;\;\;\;\;\;\;
\;\;\;\;\;\;\;\;\;\;\;\;\; && 
t_{32} \;=\; (\kappa_0 - c_v) y_\eta \;\;\;\;\;\;\;
\;\;\;\;\;\;\;\;\;\;\;\;\;\;\;\;\;\;\;\;\;\;
t_{33} \;=\; 1+3\kappa_0 c_v \nonumber 
\eea
where the $x$'s, $y$'s and $\kappa$'s are defined in equations
(\ref{k0})~and~(\ref{xys}).

It is convenient to make an expansion in
powers of $\epsilon = v/f$:
\bea
T \; =\; {\bf 1} \;+\; T_2 \frac{\epsilon^2}{12} \;+\;
T_4 \frac{\epsilon^4}{6}
\;+\; \cdot \cdot \;\;\;\;\;\;\;\;\;\;\;\;\;\;\;\;\;\;\;\;\;\;\;\;
\nonumber \\
T_2 = 
\left(\begin{array}{ccc}
1/2 & - 2.183 \sqrt{5} & 2\sqrt{2} \\
\frac{2.183}{2}\sqrt{5} &  5/2 & -2\sqrt{10}  \\
-\sqrt{2} & \sqrt{10}  & 1
\end{array}\right)
\;\;\;
T_4 = 
\left(\begin{array}{ccc}
-0.5682 & 0.6200  & \frac{41}{30\sq2} \\
0.1315 &  -0.6093 & \frac{-41}{6\sqrt{10}}  \\
-0.1288 & 1.6184  & \frac{-17}{240}
\end{array}\right)
\eea
where we have taken $t_w=0.536$, thus
$(10+t^2_w)/(5-t^2_w) \,=2.183$.

In terms of the new (mass eigenstates) fields the neutral boson
Lagrangian becomes:
\bea
{\cal L}_{\rm kin} &=& 
 \frac{1}{2} (\partial^\mu h)^2 \; +\;
\frac{1}{2} (\partial^\mu \pi^{\prime 0})^2 \; +\;
M_Z Z^{\mu} \partial_\mu \pi^{\prime 0}
\nonumber \\
&+& 
\frac{1}{2} (\partial^\mu \omega^{\prime 0})^2 \; +\;
\frac{1}{2} (\partial^\mu \eta^{\prime})^2 \; +\;
\frac{1}{2} (\partial^\mu \phi_P^{\prime 0})^2 \; +\;
\frac{1}{2} (\partial^\mu \phi^{\prime 0})^2 
\nonumber \\
&+& M_{Z_H} Z_H^{\mu} \partial_\mu \omega^{\prime 0}\; +\;
 M_{A_H} A_H^{\mu} \partial_\mu \eta^{\prime}
\eea
where the mixing terms like $Z^{\mu} \partial_\mu \pi^{\prime 0}$
are canceled (after integration by parts) when we add
the usual gauge-fixing terms:
\bea
\Delta {\cal L} &=&
\frac{-1}{2\xi_A} \left( \partial_\mu A^{\mu} \right)^2 \;-\;
\frac{1}{2\xi_Z} \left( \partial_\mu Z^{\mu} - M_Z \xi_Z 
\pi^{\prime 0} \right)^2 \nonumber \\
&-&
\frac{1}{2\xi_{A_H}} \left( \partial_\mu A_H^{\mu} - 
M_{A_H} \xi_{A_H} \eta^{\prime} \right)^2 \;-\;
\frac{1}{2\xi_{Z_H}} \left( \partial_\mu Z_H^{\mu} - 
M_{Z_H} \xi_{Z_H} \omega^{\prime 0} \right)^2 
\eea


\section{Feynman rules.}
We want to show the Feynamn rules that we used.  The scalar fields
are not the original interaction eigenstates but the mass eigenstates
that are written as $\pi^{\prime \pm}$, $\pi^{\prime 0}$, etc.
We have dropped the $\prime$ symbol to simplify the notation.
Table~\ref{wbosonic} shows bosonic vertices that involve one
charged $W^-$ SM boson.  Tables~\ref{chargedscalar},
\ref{SMneutralscalar} and \ref{oddneutralscalar} show vertices
for fermions and charged, T-even neutral and T-odd neutral
scalar bosons respectively.
For the fermion-gauge boson interactions we refer the reader to
tables V and VI of Ref.~\cite{belyaev-chen}.  We have carefully
verified that our rules agree with the ones there.
We have written in Table~\ref{fermion-gaugeboson} some others that
do not appear in \cite{belyaev-chen}.  Other types of interactions,
like four-boson vertices or dimension-5 $\bar f f \phi \phi$
($\phi$ any scalar) vertices can be found in Ref.~\cite{hubisz-meade}.

Please notice the definitions
($\epsilon = v/f$, $r=\lambda_1/\lambda_2$):
\begin{eqnarray*}
S_{\lambda \nu \mu} &\equiv& (p_{-}-p_{+})_\lambda g_{\mu \nu} \, +
\, (p_{+}-p_{0})_\mu g_{\nu \lambda} \, + \, (p_{0}-p_{-})_\nu
g_{\lambda \mu} \, \\
q_w&\equiv&\frac{10+t_w^2}{5-t_w^2}
\end{eqnarray*}

\begin{table}[h]

\begin{tabular}{|lc||lc|}
\hline Interaction $\;\;\;\;$  &  Feynman rule  & Interaction
$\;\;\;\;$  &   Feynman rule \tabularnewline \hline \hline
$W_\mu^-W_\nu^+ A_\lambda$ & $i g\, s_w\,S_{\lambda \nu \mu}$ &
$W_\mu^- W_{H\nu}^+ \phi^0_P$ & $ \sim \epsilon^4$ \tabularnewline \hline
$W_\mu^- W_\nu^+ Z_\lambda$ &$i g\, c_w\,S_{\lambda \nu \mu}$ &
$W_\mu^- \omega^+ A_{H\nu}$ & $i
g^2\,f\,t_w(\frac{5}{20-4t_W^2}-\frac{1}{4}) \epsilon^2\,g_{\mu\nu}$
\tabularnewline\hline
$W_\mu^- \pi^+ A_\nu$ &$i g\,s_w
M_W(1-\frac{1}{12}\epsilon^2)g_{\mu\nu}$ &
$W_\mu^- \omega^+ Z_{H\nu}$ & $-i g^2\,f(1-\frac{3}{8}\epsilon^2)g_{\mu\nu}$ \tabularnewline \hline
$W_\mu^- \pi^+ Z_\nu$ & $-i g\,s_w^2 M_Z(1-\frac{1}{12}\epsilon^2)g_{\mu\nu}$  &
$W_\mu^- \phi^+ A_{H\nu}$ & $ g^2\frac{t_w}{4}f \epsilon^2 g_{\mu\nu}$ \tabularnewline \hline 
$W_\mu^- W_\nu^+ h$ & $i g\,M_W(1-\frac{1}{3}\epsilon^2)g_{\mu\nu}$ &
$W_\mu^- \phi^+ Z_{H\nu}$ & $i g^2\,\frac{f}{4}(\frac{2}{3}+\frac{5}{3}i)\epsilon^2g_{\mu\nu}$
\tabularnewline \hline 
$W_\mu^- W_\nu^+ \pi^0$ & $ 0 $&
$W_\mu^- \omega^+ \omega^0$ & $
g(1-\frac{1}{8}\epsilon^2)(p_{\omega^+\mu}-p_{0\mu})$ 
\tabularnewline\hline 
$W_\mu^- \pi^+ h$ & $i \frac{g}{2}[p_{\pi\mu}(1-\frac{5}{12}\epsilon^2)-p_{h\mu}(1-\frac{1}{12}\epsilon^2)]$ 
& $W_\mu^-
\omega^+ \eta$ &\hspace{-0.3cm}
$g\frac{\sqrt{5}}{12}\epsilon^2[(q_w-\frac{3}{2})
p_{\eta\mu}-(q_w+\frac{1}{2})p_{\omega^+\mu}]$\tabularnewline \hline
$W_\mu^- \pi^+ \pi^0$ & $\frac{g}{2}(p_{\pi^+\mu}-p_{0\mu})$ &
$W_\mu^- \omega^+ \phi^0$ & $-\frac{g}{12\sqrt{2}}\epsilon^2[(1+2i)p_{\phi \mu}-p_{\omega
\mu}]$ \tabularnewline \hline 
$W_\mu^- W_{H\nu}^+ A_{H\lambda}$ & $-i g\,s_h\,S_{\lambda\nu\mu}$ & $W_\mu^- \omega^+ \phi^0_P$ & $i
g\sqrt{2}\frac{1}{12}\epsilon^2[\frac{1}{2}p_{\phi\mu}-(\frac{1}{2}+i)p_{\omega\mu}]$
\tabularnewline \hline 
$W_\mu^- W_{H\nu}^+ Z_{H\lambda}$ & $i g\,c_h\, S_{\lambda\nu\mu}$ & $W_\mu^- \phi^+ \omega^0$ &
$-\frac{g}{6}(1+\frac{7}{4}i)(p_{\phi^+\mu}-p_{\omega_0\mu})\epsilon^2$
\tabularnewline \hline 
$W_\mu^- W_{H\nu}^+ \eta$ & $\frac{\sqrt{5}}{2}g^2 f(1+q_w)\epsilon^2\,g_{\mu\nu} $ & $W_\mu^-
\phi^+ \eta$ & $-i g\frac{\sqrt{5}}{12}(\frac{5}{2}p_{\eta\mu}-\frac{1}{2}p_{\phi^+\mu})\epsilon^2$
\tabularnewline \hline 
$W_\mu^- W_{H\nu}^+ \phi^0$ & 0 &
$W_\mu^- \phi^+ \phi^0$ & $\frac{g}{\sqrt{2}}[p_{0\mu}-p_{\phi^+\mu}-\frac{1}{24}\epsilon^2(5p_{0\mu}-p_{\phi^+\mu})] $
\tabularnewline \hline 
$W_\mu^- W_{H\nu}^+ \omega^0$ & $g^2 f(1-\frac{3}{8}\epsilon^2) g_{\mu \nu}$  & $W_\mu^- \phi^+ \phi^0_P$ & \hspace{-0.4cm}$-i
\frac{g}{\sqrt{2}}[p_{\phi^0\mu}-p_{\phi^+\mu}-\frac{1}{24}(5p_{\phi^0\mu}-p_{\phi^+\mu})\epsilon^2]$
\tabularnewline \hline
\end{tabular}
\caption{Feynman rules for three boson $W^-$ vertices.  Some of these rules
also appear in Ref.~\cite{buras2}.  Here, $M_W$ and $M_Z$ stand for the SM
mass of the $W^\pm$ and $Z$ bosons.
\label{wbosonic}}
\end{table}

\begin{eqnarray*}
A_R&\equiv&i\sqrt{2}\lambda_d\left[ -1+\frac{r^4}{2(1+r^2)^2}\epsilon^2\right]\\
A_L&\equiv&i\frac{\sqrt{2}\lambda_2 r}{\sqrt{1+r^2}}\left[1-\frac{1+3r^4}{4(1+r^2)^2}\epsilon^2\right]\\
B_R&\equiv&i\frac{\sqrt{2}\lambda_d\, r^2}{1+r^2}\left[ -\epsilon + \frac{5-2r^2+2r^4}{6(1+r^2)^2}\epsilon^3\right]\\
B_L&\equiv&i\sqrt{2}\lambda_2\frac{r^2}{\sqrt{1+r^2}}\left[1-\frac{\epsilon^2}{4}\frac{3+r^4}{(1+r^2)^2}\right]\\
D_R &\equiv&(i+4)\kappa \frac{\epsilon^2}{24}\\
D_L&\equiv&  \frac{i r\lambda_2}{\sqrt{2}\sqrt{1+r^2}}
\left[-\epsilon+\frac{\epsilon^3}{24}\frac{7-4i+(2-8i)r^2+(19-4i)r^4}{(1+r^2)^2}\right]\\
E_L&\equiv&\frac{i r^2 \lambda_2}{\sqrt{2(1+r^2)}}\left[ -\epsilon + \frac{19+4i+(2+8i)r^2+(7+4i)r^4}{24(1+r^2)^2}\epsilon^3  \right]\\
a_R&\equiv&-\kappa \left[1-\frac{r^4\epsilon^2}{2(1+r^2)^2}\right]\\
a_L&\equiv&\frac{r\lambda_2}{\sqrt{2}\sqrt{1+r^2}}\left(\epsilon-\frac{i\epsilon^3
[2-7i+(4-2i)r^2+(2-19i)r^4]}{24(1+r^2)^2}\right)\\
b_R&\equiv&\frac{r^2 \kappa}{1+r^2}\left[ -\epsilon+\frac{5-2r^2+2r^4}{6(1+r^2)^2}\epsilon^3 \right] \\
b_L&\equiv&\frac{r^2 \lambda_2}{\sqrt{2(1+r^2)}}\left[ \epsilon-\frac{19+2i+ (2+4i)r^2 + (7+2i)r^4}{24(1+r^2)^2}\epsilon^3 \right]
\end{eqnarray*}
\begin{table}[h]
\begin{tabular}{|cc||cc|}
\hline Interaction $\;\;\;\;$  &  Feynman rule  & Interaction
$\;\;\;\;$  &   Feynman rule \tabularnewline \hline \hline 
$\bar t^+b^+ \pi^+$ &$A_R P_R + A_L P_L$& 
$\bar t^- b^+ \omega^+$ & $i\frac{\lambda_d}{\sqrt{2}}[-\epsilon+(1-2i)\frac{1}{24}\epsilon^3] P_R + i \kappa P_L$ \tabularnewline \hline 
$\bar T^+ b^+ \pi^+$ & $ B_R P_R + B_L P_L$& 
$\bar T^- b^+ \omega^+$ & $0$  \tabularnewline \hline 
$\bar t^- b^- \pi^+$ & $ -i\frac{\kappa}{4}(\epsilon+\frac{1}{24}\epsilon^3)\gamma_5$ & $\bar t^+ b^- \phi^+$ &
$D_R\,P_R+D_L\,P_L$ \tabularnewline \hline 
$\bar T^-b^- \pi^+$ & $ 0 $& 
$\bar T^+ b^- \phi^+$ &$ i \frac{r^2 \kappa}{24(1+r^2)}[1-4i]\epsilon^3 P_R + E_L P_L $\tabularnewline \hline 
$\bar t^+ b^- \omega^+$ & $i[a_R\,P_R+a_L\,P_L]$  & 
$\bar t^- b^+ \phi^+$ & $i\frac{\lambda_d}{\sqrt{2}} [\epsilon-\frac{(1+4i)}{24}\epsilon^3] P_R -i(4i+1)\frac{\kappa}{24}\epsilon^2 P_L$ \tabularnewline \hline 
$\bar T^+ b^- \omega^+$ & $i[b_R P_R + b_L P_L]$ & 
$\bar T^- b^+ \phi^+$ & $ 0 $
\tabularnewline \hline
\end{tabular}
\caption{Feynman rules for fermion-charged scalar vertices.
\label{chargedscalar}}
\end{table}

\begin{eqnarray*}
G_R&\equiv&\frac{1}{r(1+r^2)}\left[\epsilon+\frac{3r^6-2r^4+8r^2-2}{6(1+r^2)^2}\epsilon^3\right]\\
G_L&\equiv&-1+\frac{3+r^2+r^4}{2(1+r^2)^2}\epsilon^2\\ 
J_R&\equiv&\frac{r}{1+r^2}\left[\epsilon-\frac{\epsilon^3(13-4r^2+13r^4)}{12(1+r^2)^2}\right]\\
J_L&\equiv&\frac{3\epsilon^2(1+r^4)}{4(1+r^2)^2}-1
\end{eqnarray*}
\begin{table}[h]
\begin{tabular}{|cc||cc|}
\hline Interaction $\;\;\;\;$  &  Feynman rule  & Interaction
$\;\;\;\;$  &   Feynman rule \tabularnewline \hline \hline 
$\bar t^+ t^+ h$ & $i \frac{r\lambda_2}{\sqrt{1+r^2}}(-1+\frac{2+r^2+2r^4}{2(1+r^2)^2}
\epsilon^2)$&
$\bar t^+ t^+ \pi^0$ &
$-\frac{r\lambda_2}{\sqrt{1+r^2}}[1-\frac{(1+5r^4)}{4(1+r^2)^2}\epsilon^2]\gamma_5$
\tabularnewline \hline
$\bar T^+ T^+ h$ & $i \lambda_2[\frac{r^2}{\left(r^2+1\right)^{3/2}}\epsilon+\frac{r^2 \left(3 r^6+r^4+8 r^2-5\right) }{6 \left(r^2+1\right)^{7/2}}
\epsilon^3] $&
$\bar T^+ T^+ \pi^0$ & $\frac{r^4 \lambda_2}{(1+r^2)^{\frac{3}{2}}}
[-\epsilon+\frac{19-4r^2+7r^4}{12(1+r^2)^2}\epsilon^3]\gamma_5 $
\tabularnewline \hline
$\bar T^+ t^+ h$ &
$i\frac{r^2\lambda_2}{\sqrt{1+r^2}}(G_R\,P_R+G_L\,P_L)$ &
$\bar T^+ t^+ \pi^0$ &
$-\frac{r^2\lambda_2}{\sqrt{1+r^2}}(J_R P_R+J_L P_L)$
\tabularnewline \hline
$\bar b^+ b^+ h$ & $i \lambda_d(-1+\frac{3 }{4}\epsilon ^2)$ &
$\bar b^+ b^+ \pi^0$ & $ \lambda_d \,\gamma_5$
\tabularnewline \hline
$\bar t^- t^- h$ &  $0$ &
$\bar t^- t^- \pi^0$ & $-\frac{\kappa}{2\sqrt{2}}(\epsilon+\frac{1}{24}\epsilon^3)\gamma_5 $
\tabularnewline \hline
$\bar T^- T^- h$ & $ 0 $&
$\bar T^- T^- \pi^0$ & $ 0 $
\tabularnewline \hline
$\bar T^- t^- h$ & $ 0 $ &
$\bar T^- t^- \pi^0$ & $ 0 $
\tabularnewline \hline
$\bar b^- b^- h$ & $0$ &
$\bar b^- b^- \pi^0$ & $0$
\tabularnewline \hline
\end{tabular}
\caption{Feynman rules for fermion-neutral T-even scalar vertices.
\label{SMneutralscalar}}
\end{table}

\begin{eqnarray*}
K_R&\equiv&-\frac{\kappa}{\sqrt{2}}\left[1-\left(1+q_w+\frac{12r^4}{(1+r^2)^2}\right)\frac{\epsilon^2}{24}\right]\\
K_L&\equiv&\frac{r\lambda_2}{2\sqrt{1+r^2}}\left[ \epsilon-\left(q_w+\frac{3-6r^2+15r^4}{(1+r^2)^2}\right)\frac{\epsilon^3}{24}  \right]\\
M_R &\equiv& \frac{r^2 \kappa}{\sqrt{2}(1+r^2)}\left(-\epsilon + \left[q_w+\frac{21-6r^2+9r^4}{(1+r^2)^2}\right]\frac{1}{24}\epsilon^3\right)\\
M_L &\equiv&  \frac{r^2 \lambda_2}{2\sqrt{1+r^2}}\left(\epsilon - \left[q_w+\frac{15-6r^2+3r^4}{(1+r^2)^2}\right]\frac{1}{24}\epsilon^3\right)\\
N_R&\equiv&\frac{\kappa}{\sqrt{10}}\left[
1+\left(10\,q_w-5-\frac{12r^4}{(1+r^2)^2}\right)\frac{\epsilon^2}{24} \right]\\
N_L&\equiv&-\frac{r\lambda_2}{2\sqrt{5(1+r^2)}}\left[
\epsilon+\left(10\,q_w+\frac{17+46r^2+5r^4}{(1+r^2)^2}\right)\frac{\epsilon^3}{24}
\right]\\
V_L&\equiv&-\frac{r\lambda_2}{\sqrt{2(1+r^2)}}\left[\epsilon-\frac{\epsilon^3(7+8r^2+13r^4)}{12(1+r^2)^2}\right]\\
\Xi_R&\equiv&\frac{r^2 \kappa}{\sqrt{10}(1+r^2)}\left[ \epsilon + \left(10\, q_w -\frac{25+2r^2 +13 r^4}{(1+r^2)^2}\right)\frac{\epsilon^3}{24}\right]\\
\Xi_L&\equiv&-\frac{r^2 \lambda_2}{2\sqrt{5(1+r^2)}}\left[ \epsilon + \left( 10\, q_w + \frac{5+46r^2+17r^4}{(1+r^2)^2} \right)\frac{\epsilon^3}{24} \right]\\
\beta_R&\equiv&-\frac{r^2\kappa}{4(1+r^2)^2}\epsilon^3\\
\beta_L&\equiv&\frac{r^2\lambda_2}{\sqrt{2(1+r^2)}}\left[ -\epsilon + \frac{13+8r^2+7r^4}{12(1+r^2)^2}\epsilon^3  \right]
\end{eqnarray*}

\begin{table}[h]
\begin{tabular}{|cc||cc|}
\hline Interaction $\;\;\;\;$  &  Feynman rule  & Interaction
$\;\;\;\;$  &   Feynman rule \tabularnewline \hline \hline
$\bar t^+ t^- \omega^0 $ & $K_R\,P_R+K_L\,P_L$&
$\bar t^+ t^- \phi^0$ & $\frac{i r\lambda_2}{\sqrt{2(1+r^2)}}\left[\epsilon-\frac{1-4r^2+7r^4}{12(1+r^2)^2}\epsilon^3\right]P_L$
\tabularnewline \hline
$\bar T^+ t^- \omega^0$ & $ M_R P_R + M_L P_L$ &
$\bar T^+ t^- \phi^0$ & $ i\frac{r^2 \lambda_2}{\sqrt{2(1+r^2)}}\left[\epsilon +\frac{-7+4r^2-r^4}{12(1+r^2)^2}\epsilon^3 \right]P_L$
 \tabularnewline \hline
$\bar T^- t^+ \omega^0$ & $ \frac{r \lambda_2}{12\sqrt{1+r^2}}(2-q_w)\epsilon^2 P_R$ &
$\bar T^- t^+ \phi^0$ & $ 0 $
\tabularnewline \hline
$\bar T^+ T^- \omega^0$ & $\frac{q_w-2}{12\sqrt{1+r^2}}r^2 \lambda_2\,\epsilon^2 P_L$&
$\bar T^+ T^- \phi^0$ & $ 0 $ \tabularnewline \hline
$\bar b^+ b^- \omega^0$ &
$\frac{\kappa}{\sqrt{2}}\left[1+\frac{1+q_w}{24}\epsilon^2\right]P_R$&
$\bar b^+ b^- \phi^0$ & $ 0 $
\tabularnewline \hline
$\bar t^+ t^- \eta$ & $N_R\,P_R+N_L\,P_L $ &

$\bar t^+ t^- \phi^0_P$ &
$-\frac{\kappa\epsilon^2}{4}P_R\,+\,V_L\,P_L$
\tabularnewline \hline
$\bar T^+ t^- \eta$ & $\Xi_R P_R + \Xi_L P_L $ &
$\bar T^+ t^- \phi^0_P$ & $\beta_R P_R + \beta_L P_L $
\tabularnewline \hline
$\bar T^- t^+ \eta$ &
$-\frac{2r\lambda_2}{\sqrt{5(1+r^2)}}\left[1+\frac{1+6r^2-3r^4}{8(1+r^2)^2}\epsilon^2\right]P_R
$ &
$\bar T^- t^+ \phi^0_P$ & $ \sim \epsilon^4$
\tabularnewline \hline
$\bar T^+ T^- \eta$ & $\frac{2r^2 \lambda_2}{\sqrt{5(1+r^2)}}[1+\frac{r^4+6r^2-3}{8(1+r^2)^2}\epsilon^2]P_L $ &
$\bar T^+ T^- \phi^0_P$ & $ 0 $
\tabularnewline \hline
$\bar b^+ b^- \eta$ & $\frac{\kappa}{\sqrt{10}}\left[
1-\frac{5(2q_w-1)}{24}\epsilon^2 \right]P_R $ &
$\bar b^+ b^- \phi^0_P$ & $i\frac{\lambda_d}{\sqrt{2}}(\frac{1}{3}\epsilon^3-\epsilon) $
\tabularnewline \hline
\end{tabular}
\caption{Feynman rules for fermion-neutral T-odd scalar vertices.
\label{oddneutralscalar}}
\end{table}

\begin{eqnarray*}
 C_t& \equiv&c_L^2\left(c_w^2-\frac{1}{3}s_w^2\right) -\frac{4}{3}s^2_L s_w^2\\
C_T& \equiv&s_L^2\left(c_w^2-\frac{1}{3}s_w^2\right) -\frac{4}{3}c^2_L s_w^2
\end{eqnarray*}

\begin{table}
\begin{tabular}{|cc||cc|}
\hline Interaction $\;\;\;\;$  &  Feynman rule  & Interaction
$\;\;\;\;$  &   Feynman rule \tabularnewline \hline \hline
$\bar b^- t^- W^-_\mu $ & $i\frac{g}{\sqrt{2}}\gamma_\mu $&
$\bar t^+ t^+ Z_\mu$ & $i\frac{g}{2 c_w}\gamma_\mu [C_t P_L-\frac{4}{3}s_w^2 P_R]$
\tabularnewline \hline
$\bar b^- T^- W^-_\mu$ & $0 $ &
$\bar T^+ T^+ Z_\mu$ & $ i\frac{g}{2 c_w}\gamma_\mu [C_T P_L-\frac{4}{3}s_w^2 P_R]$
 \tabularnewline \hline
$\bar b^+ T^- W^-_{H\mu}$ & $ 0$ &
$\bar b^+ b^+ Z_\mu$ & $ i\frac{g}{2c_w}\gamma_\mu[\frac{2}{3}s_w^2 P_R + (\frac{2}{3}s_w^2-1)P_L] $
\tabularnewline \hline
$\bar t^{\pm} t^{\pm} A_\mu$ & $i\frac{2}{3} e\gamma_\mu$&
$\bar t^- t^- Z_\mu$ & $i\frac{g}{2c_w}\gamma_\mu[c_w^2-\frac{1}{3}s_w^2] $ 
\tabularnewline \hline
$\bar T^{\pm} T^{\pm} A_\mu$ &$i\frac{2}{3} e\gamma_\mu $&
$\bar T^- T^- Z_\mu$ & $ -i\frac{2g}{3c_w}s_w^2 \gamma_\mu $
\tabularnewline \hline
$\bar b^{\pm} b^{\pm} A_\mu$ & $-i\frac{1}{3}e \gamma_\mu $ &
$\bar T^- t^- Z_\mu$ &$0 $
\tabularnewline \hline
$\bar T^{+} t^{+} Z_\mu$ & $i\frac{g}{2 c_w}c_L s_L \gamma_\mu P_L $ &
$\bar b^- b^- Z_\mu$ &$i\frac{g}{2 c_w}\gamma_\mu[\frac{2}{3}s_w^2-1] $
\tabularnewline \hline
\end{tabular}
\caption{Some Feynman rules for fermion-gauge boson vertices.
All the other rules appear in Ref.~\cite{belyaev-chen} and in
Ref.~\cite{buras2}.
\label{fermion-gaugeboson}}
\end{table}





\begin{thebibliography}{1}


\bibitem{topreviews}
  T.~Han, {\it The 'Top Priority' at the LHC},
 arXiv:0804.3178 [hep-ph];
  T.~M.~P.~Tait,
  Nucl.\ Phys.\ Proc.\ Suppl.\  {\bf 177-178}, 11 (2008);
A. Quadt, {\it Top quark physics at hadron colliders},
\EPC {\bf 48} (2006) 835; 
D. Chakraborty, J. Konigsberg, D. Rainwater,
Annu. Rev. Part. Nucl. Sci. {\bf 53} (2003) 301;
M. Beneke, et.al.,
{\it Top Quark Physics: 1999 CERN Workshop on the SM Physics
(and more) at the LHC} (hep-ph/0003033);
  F.~Larios, R.~Martinez and M.~A.~Perez,
  Int.\ J.\ Mod.\ Phys.\  A {\bf 21}, 3473 (2006);
  R.~Martinez, M.~A.~Perez and N.~Poveda,
  Eur.\ Phys.\ J.\  C {\bf 53}, 221 (2008);
  J.~L.~Diaz-Cruz, M.~A.~Perez and J.~J.~Toscano,
  Phys.\ Lett.\  B {\bf 398}, 347 (1997).





\bibitem{singletops}
  T.~M.~P.~Tait,
  Phys.\ Rev.\  D {\bf 61}, 034001 (1999);
  T.~M.~P.~Tait and C.~P.~Yuan,
  Phys.\ Rev.\  D {\bf 63}, 014018 (2000);
  Q.~H.~Cao, J.~Wudka and C.~P.~Yuan,
  Phys.\ Lett.\  B {\bf 658}, 50 (2007);
  Q.~H.~Cao and J.~Wudka,
  Phys.\ Rev.\  D {\bf 74}, 094015 (2006);
  F.~Larios, M.~A.~Perez and C.~P.~Yuan,
  Phys.\ Lett.\  B {\bf 457}, 334 (1999);
  U.~Baur, A.~Juste, L.~H.~Orr and D.~Rainwater,
  Nucl.\ Phys.\ Proc.\ Suppl.\  {\bf 160}, 17 (2006);
  B.~Grzadkowski and M.~Misiak,
   Phys.\ Rev.\  D {\bf 78}, 077501 (2008);
arXiv:0802.1413 [hep-ph];
  B.~Sahin and I.~Sahin,
  Eur.\ Phys.\ J.\  C {\bf 54}, 435 (2008).


\bibitem{tevsingle}
  R.~Schwienhorst  [D0 Collaboration and CDF Collaboration],
  arXiv:0805.2175 [hep-ex];
  V.~M.~Abazov {\it et al.}  [D0 Collaboration],
  arXiv:0807.1692 [hep-ex];
  V.~M.~Abazov {\it et al.}  [D0 Collaboration],
  Phys.\ Rev.\ Lett.\  {\bf 100}, 062004 (2008);
  C.~I.~Ciobanu  [CDF Collaboration and D0 Collaboration],
  arXiv:0809.2173 [hep-ex].


\bibitem{aguilartbw}
  J.~A.~Aguilar-Saavedra, J.~Carvalho, N.~Castro, A.~Onofre and F.~Veloso,
  Eur.\ Phys.\ J.\  C {\bf 53}, 689 (2008);
  J.~A.~Aguilar-Saavedra,
  Nucl.\ Phys.\  B {\bf 804}, 160 (2008);
  F.~Hubaut, E.~Monnier, P.~Pralavorio, K.~Smolek and V.~Simak,
  Eur.\ Phys.\ J.\  C {\bf 44S2}, 13 (2005).


\bibitem{oakes}J. Cao, R.J. Oakes, F. Wang and J.M. Yang,
\PRD {\bf 68} (2003) 054019;
  M.~Beccaria, C.~M.~Carloni Calame, G.~Macorini, E.~Mirabella, 
F.~Piccinini, F.~M.~Renard and C.~Verzegnassi,
  Phys.\ Rev.\  D {\bf 77}, 113018 (2008).


\bibitem{qiao}X. Wang, Q. Zhang and Q. Qiao,
\PRD {\bf 71} (2005) 014035;
  X.~Wang, Y.~Xi, Y.~Zhang and H.~Jin,
  Phys.\ Rev.\  D {\bf 77}, 115006 (2008).


\bibitem{chivukulalecture}
S.~Chivukula, {\it Models of Electroweak Symmetry Breaking},
hep-ph/9803219.


\bibitem{arkani01}
  N.~Arkani-Hamed, A.~G.~Cohen, and H.~Georgi,
\PLB {\bf 513} (2001) 232;
  N.~Arkani-Hamed, A.~G.~Cohen, T.~Gregoire and J.~G.~Wacker,
  JHEP {\bf 0208}, 020 (2002);
  N.~Arkani-Hamed, A.~G.~Cohen, E.~Katz, A.~E.~Nelson, 
T.~Gregoire and J.~G.~Wacker,
  JHEP {\bf 0208}, 021 (2002);
  W.~Skiba and J.~Terning,
  Phys.\ Rev.\  D {\bf 68}, 075001 (2003).


\bibitem{arkani02}
  N.~Arkani-Hamed, A.~G.~Cohen, E.~Katz and A.~E.~Nelson,
  JHEP {\bf 0207}, 034 (2002).


\bibitem{littlereviews}
  M.~Perelstein,
  Prog.\ Part.\ Nucl.\ Phys.\  {\bf 58}, 247 (2007)
  M.~Schmaltz and D.~Tucker-Smith,
  Ann.\ Rev.\ Nucl.\ Part.\ Sci.\  {\bf 55}, 229 (2005)



\bibitem{han-logan}
T.~Han, H.~E.~Logan, B.~McElrath and L.~T.~Wang,
\PRD {\bf 67}, 095004 (2003);
  G.~Burdman, M.~Perelstein and A.~Pierce,
  Phys.\ Rev.\ Lett.\  {\bf 90}, 241802 (2003)
  [Erratum-ibid.\  {\bf 92}, 049903 (2004)];
  T.~Han, H.~E.~Logan and L.~T.~Wang,
  JHEP {\bf 0601}, 099 (2006);
  M.~Perelstein, M.~E.~Peskin and A.~Pierce,
  Phys.\ Rev.\  D {\bf 69}, 075002 (2004);
  D.~E.~Kaplan, M.~Schmaltz and W.~Skiba,
  Phys.\ Rev.\  D {\bf 70}, 075009 (2004);
  G.~Azuelos {\it et al.},
  Eur.\ Phys.\ J.\  C {\bf 39S2}, 13 (2005);
  J.~A.~Conley, J.~L.~Hewett and M.~P.~Le,
  Phys.\ Rev.\  D {\bf 72}, 115014 (2005);
  C.~X.~Yue, L.~Zhou and S.~Yang,
  Eur.\ Phys.\ J.\  C {\bf 48}, 243 (2006);
  C.~X.~Yue, S.~Yang and L.~H.~Wang,
  Europhys.\ Lett.\  {\bf 76}, 381 (2006);
  Y.~B.~Liu, J.~F.~Shen and X.~L.~Wang,
  arXiv:hep-ph/0610350;
  L.~Wang, W.~Wang, J.~M.~Yang and H.~Zhang,
  Phys.\ Rev.\  D {\bf 75}, 074006 (2007);
  S.~K.~Kang, C.~S.~Kim and J.~Park,
  Phys.\ Lett.\  B {\bf 666}, 38 (2008);
  J.~Boersma,
  Phys.\ Rev.\  D {\bf 74}, 115008 (2006);
  G.~A.~Gonzalez-Sprinberg, R.~Martinez and J.~A.~Rodriguez,
  Phys.\ Rev.\  D {\bf 71}, 035003 (2005);
  W.~Kilian, D.~Rainwater and J.~Reuter,
  Phys.\ Rev.\  D {\bf 74}, 095003 (2006)
  [Erratum-ibid.\  D {\bf 74}, 099905 (2006)];
  K.~Cheung, C.~S.~Kim, K.~Y.~Lee and J.~Song,
  Phys.\ Rev.\  D {\bf 74}, 115013 (2006);
  C.~S.~Chen, K.~Cheung and T.~C.~Yuan,
  Phys.\ Lett.\  B {\bf 644}, 158 (2007);
  K.~Cheung and J.~Song,
  Phys.\ Rev.\  D {\bf 76}, 035007 (2007);
  K.~Cheung, J.~Song, P.~Tseng and Q.~S.~Yan,
  arXiv:0806.4411 [hep-ph].




\bibitem{lhlimits}
  C.~Csaki, J.~Hubisz, G.~D.~Kribs, P.~Meade and J.~Terning,
  Phys.\ Rev.\  D {\bf 67}, 115002 (2003);
  C.~Csaki, J.~Hubisz, G.~D.~Kribs, P.~Meade and J.~Terning,
  Phys.\ Rev.\  D {\bf 68}, 035009 (2003);
  J.~L.~Hewett, F.~J.~Petriello and T.~G.~Rizzo,
  JHEP {\bf 0310}, 062 (2003);
  W.~Kilian and J.~Reuter,
  Phys.\ Rev.\  D {\bf 70}, 015004 (2004);
  Z.~Han and W.~Skiba,
  Phys.\ Rev.\  D {\bf 72}, 035005 (2005);
  M.~C.~Chen and S.~Dawson,
  Phys.\ Rev.\  D {\bf 70}, 015003 (2004);
  C.~x.~Yue and W.~Wang,
  Nucl.\ Phys.\  B {\bf 683}, 48 (2004);
  M.~C.~Chen,
  Mod.\ Phys.\ Lett.\  A {\bf 21}, 621 (2006).




\bibitem{tparity}
  H.~C.~Cheng and I.~Low,
  JHEP {\bf 0408}, 061 (2004);
  H.~C.~Cheng and I.~Low,
  JHEP {\bf 0309}, 051 (2003);
  H.~C.~Cheng,
  arXiv:0710.3407 [hep-ph];
  H.~C.~Cheng, I.~Low and L.~T.~Wang,
  Phys.\ Rev.\  D {\bf 74}, 055001 (2006);
  M.~Perelstein,
  Pramana {\bf 67}, 813 (2006)
  W.~Kilian, D.~Rainwater and J.~Reuter,
  Phys.\ Rev.\  D {\bf 71}, 015008 (2005).


\bibitem{low}
 I.~Low,
  JHEP {\bf 0410}, 067 (2004).


\bibitem{hubisz-precision}
  J.~Hubisz, P.~Meade, A.~Noble and M.~Perelstein,
  JHEP {\bf 0601}, 135 (2006).


\bibitem{lhtphenomenology}
  C.~F.~Berger, M.~Perelstein and F.~Petriello,
  arXiv:hep-ph/0512053;
  M.~S.~Carena, J.~Hubisz, M.~Perelstein and P.~Verdier,
  Phys.\ Rev.\  D {\bf 75}, 091701 (2007);
  A.~Freitas and D.~Wyler,
  JHEP {\bf 0611}, 061 (2006);
  M.~M.~Nojiri and M.~Takeuchi,
  JHEP {\bf 0810}, 025 (2008);
  Y.~B.~Liu, J.~F.~Shen and X.~L.~Wang,
  arXiv:hep-ph/0610350;
  Q.~H.~Cao, C.~R.~Chen, F.~Larios and C.~P.~Yuan,
  arXiv:0801.2998 [hep-ph];
  C.~R.~Chen, K.~Tobe and C.~P.~Yuan,
\PLB {\bf 640}, 263 (2006);
  Q.~H.~Cao and C.~R.~Chen,
 \PRD {\bf 76}, 075007 (2007);
  K.~Hsieh and C.~P.~Yuan,
  Phys.\ Rev.\  D {\bf 78}, 053006 (2008);
  V.~Barger, W.~Y.~Keung and Y.~Gao,
  Phys.\ Lett.\  B {\bf 655}, 228 (2007);
  A.~Datta, P.~Dey, S.~K.~Gupta, B.~Mukhopadhyaya and A.~Nyffeler,
  Phys.\ Lett.\  B {\bf 659}, 308 (2008).

\bibitem{hill}
  C.~T.~Hill and R.~J.~Hill,
  Phys.\ Rev.\  D {\bf 76}, 115014 (2007); ibid.
  Phys.\ Rev.\  D {\bf 75}, 115009 (2007).


\bibitem{grinstein}
  B.~Grinstein and M.~Trott,
  arXiv:0808.2814 [hep-ph].

\bibitem{barcelo}
  R.~Barcelo, M.~Masip and M.~Moreno-Torres,
  Nucl.\ Phys.\  B {\bf 782}, 159 (2007).



\bibitem{belyaev-chen}
  A.~Belyaev, Chuan-Ren Chen, K. Tobe and C.-P. Yuan,
 \PRD {\bf 74}, 115020 (2006).


\bibitem{hubisz-meade}
  J.~Hubisz and P.~Meade,
 \PRD {\bf 71}, 035016 (2005).



\bibitem{tobe-matsumoto}
  S.~Matsumoto, T.~Moroi and K.~Tobe,
  Phys.\ Rev.\  D {\bf 78}, 055018 (2008).



\bibitem{buras1}
  J.~Hubisz, S.~J.~Lee and G.~Paz,
  JHEP {\bf 0606}, 041 (2006);
M. Blanke, A. Buras, A. Poschienrieder, C. Tarantino,
S. Uhlig and A. Weiler.
  JHEP {\bf 12}, 003 (2006).



\bibitem{chen-tpol}
  C.~R.~Chen, F.~Larios and C.~P.~Yuan,
\PLB {\bf 631}, 126 (2005).


\bibitem{kane}G.L. Kane, G.A. Ladinsky and C.-P. Yuan.
\Journal{\PRD}{45}{124}{1992}.


\bibitem{aguila}F. del Aguila and J.A. Aguilar-Saavedra,
\Journal{\PRD}{67}{014009}{2003}.



\bibitem{bernreuther}
  W.~Bernreuther,
  J.\ Phys.\ G {\bf 35}, 083001 (2008)
  [arXiv:0805.1333 [hep-ph]].


\bibitem{cao-li}
  Q.~H.~Cao, C.~S.~Li and C.~P.~Yuan,
  Phys.\ Lett.\  B {\bf 668}, 24 (2008).


\bibitem{foot1}
We would like to refer the
reader to the discussion in Ref.\cite{hubisz-precision}
concerning the electroweak precision constraints on LHT
and the effects of higher dimesion operators.


\bibitem{pdf} J. Pumplin, D.R. Stump, J. Huston, H.L. Lai,
P. Nadolsky and W.K. Tung, JHEP {\bf 0207} (2002) 012.


\bibitem{korner}H.S. Do, S. Groote, J.G. Korner and
M.C. Mauser, \Journal{\PRD}{67}{091501}{2003},
and references therein.
  G.~Calderon and G.~Lopez Castro,
Int. Journal Mod. Phys. A23, 3525 (2008).


\bibitem{buras2}
M. Blanke, A. Buras, A. Poschienrieder, C. Tarantino,
S. Recksiegel, S. Uhlig and A. Weiler.
  JHEP {\bf 01}, 066 (2007).


\bibitem{hollik}
  W.~Hollik, ``Renormalization of the Standard Model.''
MPI-PH-93-21, BI-TP-93-16, Apr 1993. 79pp.  See also,
T.-P. Cheng and L.-F. Li, ``Gauge Theory of elementary
particle physics'', Clarendon Press, Oxford 1984. 



\end{thebibliography}
\end{document}